\documentclass[10pt,journal,compsoc]{IEEEtran}

\usepackage{booktabs}
\usepackage{threeparttable}
\usepackage{amsmath,graphicx,cite,amssymb}
\usepackage{amsfonts,multirow,bm,array,setspace,stfloats}

\usepackage{graphicx,float,cite,amssymb}
\usepackage{graphics}
\DeclareGraphicsExtensions{.pdf,.jpeg,.png,.jpg}   
\usepackage{multirow,bm,bbm,array,setspace}
\usepackage{textcomp}
\usepackage{psfrag}
\usepackage{color}
\usepackage{pstricks,enumerate}
\usepackage{bbm}
\usepackage{subfigure}

\usepackage{algorithm,algpseudocode}
\makeatletter
\newcommand{\distas}[1]{\mathbin{\overset{#1}{\kern\z@\sim}}}%
\newsavebox{\mybox}\newsavebox{\mysim}
\newcommand{\distras}[1]{%
	\savebox{\mybox}{\hbox{\kern1pt$\scriptstyle#1$\kern1pt}}%
	\savebox{\mysim}{\hbox{$\sim$}}%
	\mathbin{\overset{#1}{\kern\z@\resizebox{\wd\mybox}{\ht\mysim}{$\sim$}}}%
}
\makeatother
\ifCLASSINFOpdf
\else
\fi
\newtheorem{proposition}{Proposition}

\newtheorem{lemma}{Lemma}

\newtheorem{example}{Example}

\newtheorem{remark}{Remark}

\begin{document}
	%\title{Full-Duplex Cellular System with Cross-Channel Transmission and Interference Cancellation}
\title{Adaptive Blind Beamforming for\\ Intelligent Surface}

\author{
\IEEEauthorblockN{
    Wenhai Lai, \IEEEmembership{Graduate Student Member,~IEEE}, Wenyu Wang, \IEEEmembership{Graduate Student Member,~IEEE},\\ Fan Xu, \IEEEmembership{Member,~IEEE}, Xin Li, Shaobo Niu, and Kaiming Shen, \IEEEmembership{Senior Member,~IEEE}
} % <-this % stops a space
\IEEEcompsocitemizethanks{
    \IEEEcompsocthanksitem Wenhai Lai, Wenyu Wang, and Kaiming Shen are with The Chinese University of Hong Kong (Shenzhen), China. E-mail: \{wenhailai, wenyuwang\}@link.cuhk.edu.cn, shenkaiming@cuhk.edu.cn.
    \IEEEcompsocthanksitem Fan Xu is with the School of Electronic Information Engineering, Tongji University, Shanghai, China. E-mail: xxiaof451@qq.com.
    \IEEEcompsocthanksitem Xin Li and Shaobo Niu are with Huawei Technologies, China. E-mail: \{razor.lixin, niushaobo\}@huawei.com.
}
\thanks{
    This work was supported in part by Guangdong Major Project of Basic and Applied
    Basic Research (No. 2023B0303000001), in part by the National Natural Science Foundation of China (NSFC) under Grant 92167202, and in part by Shenzhen Steady Funding Program.
    Wenhai Lai and Wenyu Wang made equal contributions. (Corresponding author: Kaiming Shen.)
}
}

		%Telephone: (800) 555--1212, Fax: (888) 555--1212}


\IEEEtitleabstractindextext{
\begin{abstract}
Configuring intelligent surface (IS) or passive antenna array without any channel knowledge, namely blind beamforming, is a frontier research topic in the wireless communication field. Existing methods in the previous literature for blind beamforming include the RFocus and the CSM, the effectiveness of which has been demonstrated on hardware prototypes. However, this paper points out a subtle issue with these blind beamforming algorithms: the RFocus and the CSM may fail to work in the non-line-of-sight (NLoS) channel case. To address this issue, we suggest a grouping strategy that enables adaptive blind beamforming. Specifically, the reflective elements (REs) of the IS are divided into three groups; each group is configured randomly to obtain a dataset of random samples. We then extract the statistical feature of the wireless environment from the random samples, thereby coordinating phase shifts of the IS without channel acquisition. The RE grouping plays a critical role in guaranteeing performance gain in the NLoS case. In particular, if we place all the REs in the same group, the proposed algorithm would reduce to the RFocus and the CSM.
We validate the advantage of the proposed blind beamforming algorithm in the real-world networks at 3.5 GHz aside from simulations.
\end{abstract}

}

\maketitle

\IEEEdisplaynontitleabstractindextext

\IEEEpeerreviewmaketitle

\IEEEraisesectionheading{\section{Introduction}\label{sec:introduction}}
\IEEEPARstart{P}{assive} antennas \cite{li2019towards, Arun2020RFocus}, a.k.a. reflectors \cite{han2017enhancing,xiong2017customizing}, constitute an emerging wireless technology that harnesses the existing reflected paths to enhance wireless transmission. An array of passive antennas are coordinated by choosing their {\footnotesize ON-OFF} statuses, i.e., a passive antenna reflects the incident signal if it is {\footnotesize ON} and absorbs the incident signal otherwise. The idea of passive antennas further evolves into the notion of intelligent surface (IS) \cite{ li2024secure, Zargari2023energy,shi2022intelligent,
wu2024intelligent}. An IS consists of an array of reflective elements (REs)---which induce phase shifts into their respective incident signals. To fully exploit IS or passive antennas, it entails judicious coordination of the phase shifts across the REs or the {\footnotesize ON-OFF} statuses across the passive antennas, namely \emph{passive beamforming}.

The conventional approach to the passive beamforming problem is model-driven: it first estimates the channels for all the propagation paths to formulate the problem model explicitly, and then optimizes the phase shifts (or the {\footnotesize ON-OFF} statuses) based on the problem model. However, it is difficult to carry out the model-driven approach in real-world networks because of the following two practical issues:
\begin{enumerate}[i.]
    \item Each reflected path alone is quite weak as compared to the overall channel strength and the background noise, so precise channel acquisition is technically difficult.
    \item Channel estimation for IS or passive antennas requires extra overhead and operation, which is not supported by the current network protocol.
\end{enumerate}
As such, even though the model-driven approach has been extensively studied in the literature to date, e.g., by using the semidefinite relaxation (SDR) \cite{luo2010semidefinite,wu2019intelligent, zheng2021double, zhou2020robust, zeng2020sum, xie2020max, huang2020decentralized, yao2023superimposed}, or the fractional programming (FP) \cite{shen2018fractional_p1, shen2018fractional_p2,feng2020physical,zhu2020power,shafique2020optimization,zhang2022active,zhang2021joint}, or the minorization-maximization (MM) \cite{palomar2016majorization,huang2019reconfigurable, shen2020beamforming, yu2024energy}, or the penalty convex-concave procedure (P-CCP) \cite{yao2023robust}, or the convex-hull relaxation method \cite{lai2024efficient}, yet it turns out that the existing IS prototypes \cite{Arun2020RFocus, pei2021ris, tran2020demonstration, staat2022irshield} seldom consider estimating channels.

\begin{figure}
\centering
\subfigure[LoS Case]
{
    \includegraphics[width=0.22\textwidth]{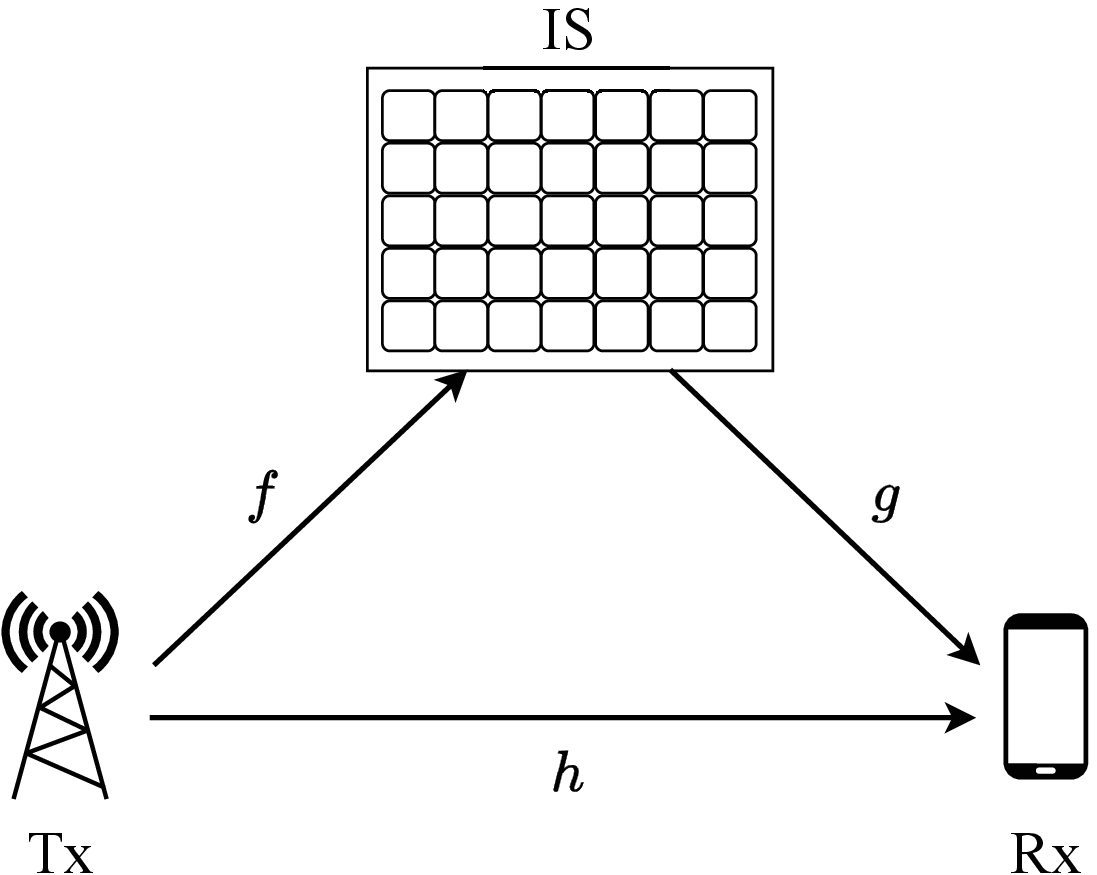}
}
\quad
\subfigure[NLoS Case]
{
    \includegraphics[width=0.22\textwidth]{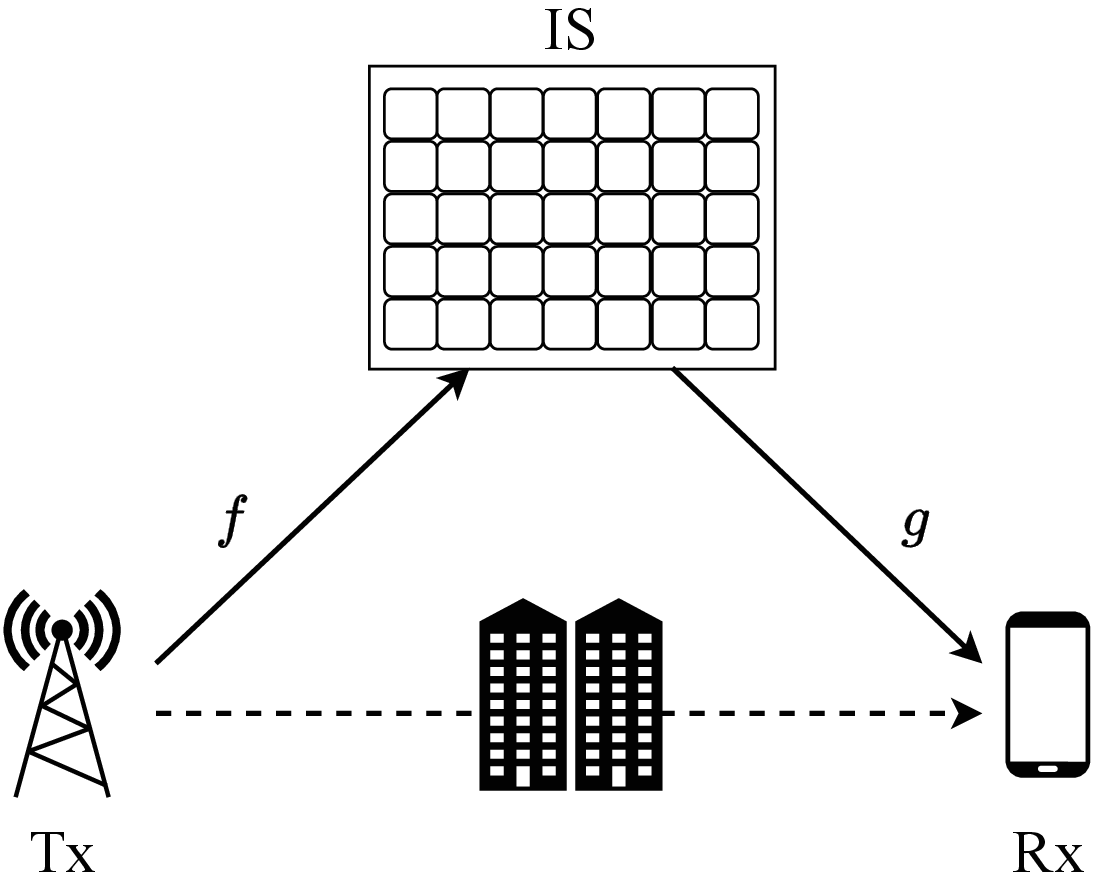}
}
\caption{Line-of-sight (LoS) case vs. non-line-of--sight (NLoS) case for the IS-assisted transmission. As a subtle issue discovered in this work, the existing blind beamforming algorithms, the RFocus \cite{Arun2020RFocus} and the CSM \cite{ren2022configuring}, can fail to work in the NLoS case.}
\label{fig:System Model}
\end{figure}

Due to the difficulties and high cost of channel acquisition, there has been an increasing interest in configuring IS without channel state information (CSI). For instance, \cite{zappone2021intelligent} examines the expected achievable rate when the phase shifts are randomly selected; \cite{psomas2021low} proposes a distributed ascent algorithm for the continuous IS beamforming based on the signal-to-noise ratio (SNR) feedback; \cite{cai2023toward} proposes an IS-aided joint index keying $M$-ary differential chaos shift keying system where the receiver can retrieve information bits by performing non-coherent correlation demodulation without CSI. The so-called beam training method \cite{you2020fast,ning2021terahertz,wang2022fast,wang2021jointbeam} constitutes another line of endeavors to perform blind beamforming, which simply tries out a sequence of possible beamformers listed in a prescribed codebook and pick the best one. Although the core idea of beam training is fairly simple, designing the codebook can be quite difficult; a common heuristic is to generate the codebook randomly. However, we will show that beam training with such a random codebook is strictly inferior to the RFocus algorithm \cite{Arun2020RFocus} and the CSM algorithm \cite{ren2022configuring}. Our work is most closely related to the above two methods which are based on statistics. Specifically, \cite{Arun2020RFocus} first proposes the idea of deciding the {\footnotesize{ON-OFF}} status for each passive antenna according to the \emph{empirical conditional average} of received signal power. The use of the empirical conditional average is further developed in \cite{ren2022configuring} to account for the phase shift optimization for IS.
More recently, the CSM algorithm is extended for multiple ISs \cite{Xu2024Coordinating} and for multiple users \cite{xu2024blind}. However, this work points out that the RFocus algorithm and the CSM algorithm are fundamentally flawed, and consequently can fail to work in the non-line-of-sight (NLoS) case wherein the direct path is much weaker than the superposition of the reflected paths, as illustrated in Fig.~\ref{fig:System Model}. To be more specific, when analyzing the performance of their algorithms, \cite{Arun2020RFocus,ren2022configuring} both implicitly assume that the direct path is sufficiently strong so that the empirical conditional average of the received signal power can accurately capture the main feature of the wireless environment. However, if the strong-direct-path assumption no longer holds (which can occur in the real world as shown in \cite{xu2023reconfiguring}), then the empirical conditional average reduces to a constant and all the subsequent steps break down. To resolve this issue, this work advocates a novel adaptive blind beamforming algorithm, which divides and groups the REs to form a virtual direct path and thereby enables the empirical conditional average-based approach.

To avoid channel estimation, some recent works \cite{2022Xuwy,2020HuangcwJSAC,2021JiangtJSAC} suggest using the deep neural network to learn the phase shift decision directly from the received pilot signals. However, its performance is sensitive to the training dataset, so we have to retrain or finetune the deep neural network whenever the wireless environment alters. Although the proposed blind beamforming method also relies on the dataset, it has much higher efficiency since only the basic operation of conditioning and averaging is required. Most importantly, blind beamforming extracts the main feature of the wireless environment by the statistic tool rather than the black-box neural network, so its performance is explainable and provable.

Further, the blind beamforming algorithm proposed in this work can be distinguished from the existing ones in that the former is examined for fading channels case whereas the latter \cite{zappone2021intelligent,psomas2021low,Arun2020RFocus,ren2022configuring,Xu2024Coordinating} mostly assume static channels. In the previous literature, the studies on passive beamforming in fading channels typically depend upon the two-state paradigm of first estimating channels and then solving the optimization problem. At the optimization stage, the optimization tools
 range from the projected gradient ascent method \cite{zhang2021large, xu2021sum, zhang2022sum} to the complex circle manifold method \cite{luo2021reconfigurable, jiang2022ris},
the parallel coordinate descent method \cite{jia2020analysis}, the genetic algorithm \cite{peng2021analysis, dai2021statistical, zhi2022power}, the particle swarm algorithm \cite{cao2022two, shekhar2022instantaneous}, and the deep learning \cite{eskandari2022statistical, ren2022long, zhongze_2022_learning}. Moreover, \cite{guo2020intelligent} suggests a model-free gradient descent method by using a large dataset of historical channel values, the main idea of which is to predict the current channel distribution from past observations.
In principle, the above methods all assume that the probability distribution of channel fading is either already known or predictable, but such statistical information for each single reflected channel can be costly to obtain when the IS comprises massive REs. In contrast, our strategy only requires the first-moment statistics (i.e., the empirical conditional average) of the received signal power.

The main contributions of the present paper are summarized in the following:
\begin{itemize}
    \item Although the previous works \cite{Arun2020RFocus,ren2022configuring} have theoretically verified the performance gain of their blind beamforming algorithms, we point out that their proofs are incomplete by showing that their algorithms can fail to work when the direct propagation path is not strong enough.
    \item To resolve the above issue, we incorporate the idea of the RE grouping into the existing empirical conditional average algorithms \cite{Arun2020RFocus,ren2022configuring}, thus allowing the blind beamforming algorithm to flexibly adapt to the current channel situation.
    \item As opposed to the existing blind beamforming studies that are limited to the static channel case, this paper proves the performance of the empirical conditional average-based approach works for fading channels.
\end{itemize}

The rest of the paper is organized as follows. Section \ref{sec:sys} describes the system model and problem formulation. Section \ref{sec:why_fail} analyzes the performance of the RFocus \cite{Arun2020RFocus} and the CSM \cite{ren2022configuring} for fading channels, and then points out that the two existing blind beamforming algorithms can fail to work in some cases. Section \ref{sec:alg_NLoS} proposes a novel RE-grouping adaptive blind beamforming algorithm; its extension to multiple users is discussed as well. Section \ref{sec:Numerical} shows the numerical results. Finally, Section \ref{sec:Conclusion} concludes the paper.

\emph{Notation:} Here and throughout, the set of complex numbers is denoted as $\mathbb{C}$. For a complex number $u, \mathfrak{Re}\{u\}$, $\mathfrak{Im}\{u\}$, $u^H$, and $\angle u$ refer to the real part, the imaginary part, the complex conjugate, and the phase of $u$, respectively. For an event $\mathcal{E}$, let $\mathbb{P}\{\mathcal{E}\}$ be its probability. Let $\mathbb{E}[X]$ be the expectation of the random variable $X$, and let $\widehat{\mathbb{E}}[X]$ be the sample mean. For a set $A$, let $|A|$ be its cardinality. The complex Gaussian distribution with mean $\mu$ and variance $\sigma^2$ is denoted as $\mathcal{CN}(\mu, \sigma^2)$. Moreover, the Bachmann-Landau notation is used extensively in the remainder of the paper. Write $f(n)=O(g(n))$ if there exists some $c>0$ such that $|f(n)| \leq c g(n)$ for $n$ sufficiently large; write $f(n)=\Omega(g(n))$ if there exists some $c>0$ such that $f(n) \geq c g(n)$ for $n$ sufficiently large; write $f(n)=\Theta(g(n))$ if $f(n)=O(g(n))$ and $f(n)=\Omega(g(n))$ both hold.

\section{System Model}
\label{sec:sys}
Consider an IS-assisted wireless communication system. The following model can be readily extended for the passive antennas. Assume that the transmitter and receiver have one antenna each, and that the IS consists of $N$ REs in total. Every RE corresponds to a reflected path from the transmitter to the receiver. Denote by $h_n\in\mathbb C$ the reflected channel associated with RE $n=1,\ldots,N$, and denote by $h_0\in\mathbb C$ the direct channel from the transmitter to the receiver. Each reflected channel $h_n$ can be factored as
\begin{equation}
\label{hn}
    h_n = f_n\times g_n,
\end{equation}
where $f_n\in\mathbb C$ is the channel from the transmitter to RE $n$ while $g_n\in\mathbb C$ is the channel from RE $n$ to the receiver. Furthermore, these channels are modeled as the Rician fading \cite{goldsmith2005wireless}:
\begin{subequations}
\label{Rician channel model}
\begin{align}
    &h_0 = \sqrt{\gamma_{00}}\left(\sqrt{\frac{\delta_{00}}{1+\delta_{00}}}\overline{h}_0+\sqrt{\frac{1}{1+\delta_{00}}}\widetilde{h}_0\right) \\
    &	f_n=\sqrt{\gamma_{0n}}\left(\sqrt{\frac{\delta_{0n}}{1+\delta_{0n}}}\overline{f}_n+\sqrt{\frac{1}{1+\delta_{0n}}}\widetilde{f}_n\right)
    \label{fn}\\
    &	g_n=\sqrt{\gamma_{n0}}\left(\sqrt{\frac{\delta_{n0}}{1+\delta_{n0}}}\overline{g}_n+\sqrt{\frac{1}{1+\delta_{n0}}}\widetilde{g}_n\right),
    \label{gn}
\end{align}
\end{subequations}
where the real scalars $\gamma_{00},\gamma_{0n},\gamma_{n0}\in[0,1]$ are the attenuation factors, the normalized complex scalars $\overline{h}_0,\overline{f}_n,\overline{g}_n$ are the line-of-sight (LoS) components, the real scalars $\delta_{00},\delta_{0n},\delta_{n0}\ge0$ are the Rician factors, and the i.i.d. standard complex Gaussian random variables  $\widetilde{h}_0,\widetilde{f}_n,\widetilde{g}_n\sim\mathcal{CN}(0,1)$ are the NLoS components. In particular, the above Rician fading model reduces to the Rayleigh fading model as the Rician factors $\delta_{00},\delta_{0n},\delta_{n0}$ tend to zero. Further, with \eqref{fn} and \eqref{gn} plugged in \eqref{hn}, each reflected channel $h_n$ can be obtained as
\begin{align}
&h_n =\sqrt{\gamma_{0n}\gamma_{n0}}\,\times\notag\\
&\left(\sqrt{\frac{\delta_{0n}\delta_{n0}}{(1+\delta_{0n})(1+\delta_{n0})}}\overline{h}_n+\sqrt{\frac{1}{(1+\delta_{0n})(1+\delta_{n0})}}\widetilde{h}_n\right),
\end{align}
where
\begin{subequations}
\begin{align}
&\overline{h}_n=\overline{f}_n\overline{g}_n\\
&\widetilde{h}_n=\overline{f}_n\widetilde{g}_n\sqrt{\delta_{0n}}+\widetilde{f}_n\overline{g}_n\sqrt{\delta_{n0}}+\widetilde{f}_n\widetilde{g}_n.
\end{align}
\end{subequations}
We remark that none of the above channel model parameters is known in our problem case. Note that the above channel model works for the active IS case \cite{wu2024intelligent} as well assuming that the amplifying parameter of IS has been fixed.

Moreover, denote by $\theta_n\in[0,2\pi)$ the phase shift of RE $n$. With the background noise $Z\sim\mathcal{CN}(0,\sigma^2)$ and the transmit signal $X\sim\mathcal{CN}(0,P)$, the received signal $Y\in\mathbb C$ is
\begin{equation}
Y = \left(h_0+\sum_{n=1}^{N}h_ne^{j\theta_n}\right)X+Z.
\end{equation}
The resulting achievable ergodic rate is given by
\begin{align}
\label{eq:ergodic_rate}
R=\mathbb{E}\left[\mathrm{log}\left(1+\frac{P}{\sigma^2}\bigg|h_0+\sum_{n=1}^{N}h_ne^{j\theta_n}\bigg|^2\right)\right],
\end{align}
where the expectation is taken over the random NLoS components $\{\widetilde{h}_0,\widetilde{f}_n,\widetilde{g}_n\}$.
But the above rate expression is difficult to tackle directly. To make it tractable, a common idea in the existing literature \cite{han2019large, jia2020analysis, gan2021ris, eskandari2022statistical, wang2021joint} is to move the expectation to the inside of log, thus obtaining an upper-bound approximation of $R$ (due to the concavity of $\log$):
\begin{equation}
\label{hat R}
\hat R=\mathrm{log}\left(1+\frac{P}{\sigma^2}\cdot\mathbb{E}\left[\bigg|h_0+\sum_{n=1}^{N}h_ne^{j\theta_n}\bigg|^2\right]\right)\ge R.
\end{equation}
Notice that maximizing the above upper bound amounts to maximizing the expectation of the overall channel power.
Further, each phase shift $\theta_n$ in practice is restricted to the discrete set
\begin{equation}
    \Phi_K=\{\omega, 2 \omega, \ldots, K \omega\},
\end{equation}
where
\begin{equation}
    \omega=\frac{2 \pi}{K},
\end{equation}
for some given positive integer $K \geq 2$. 
We seek the optimal phase shift array $\bm\theta=(\theta_1,\ldots,\theta_N)$ to maximize the approximate ergodic rate $\hat R$ in \eqref{hat R}. As a result, the passive beamforming problem can be formulated as
\begin{subequations}
\label{opt problem}
\begin{align}
    &\underset{\bm\theta}{\text{maximize}} \quad \mathbb{E}\left[\bigg|h_0+\sum_{n=1}^{N}h_ne^{j\theta_n}\bigg|^2\right]
    \label{opt problem:obj}\\
    &\text {subject to} \quad\, \theta_n \in \Phi_K,\quad n=1, \ldots, N,
    \label{opt problem:contraint}
\end{align}
\end{subequations}
where the expectation is taken over the random $\{\widetilde{h}_0,\widetilde{f}_n,\widetilde{g}_n\}$.
The difficulties of the above problem can be recognized in two respects. First, the optimizing variables $\{\theta_n\}$ are discrete. Second, the channels are unknown, i.e., $\{\theta_n\}$ must be optimized blindly.

\section{CSM Algorithm for Blind Beamforming}
\label{sec:why_fail}

The goal of this section is three-fold. First, we review the existing blind beamforming algorithms in \cite{Arun2020RFocus,ren2022configuring} for the static channels. Second, we show that these algorithms continue to work in the fading channels. Third, most importantly, we point out a flaw in the proofs of \cite{Arun2020RFocus,ren2022configuring} which can cause the failure of their algorithms in the NLoS transmission case. Although our discussion focuses on the CSM algorithm \cite{ren2022configuring}, it can be immediately extended to the RFocus algorithm \cite{Arun2020RFocus}.

\subsection{Existing Algorithms: RFocus \cite{Arun2020RFocus} and CSM \cite{ren2022configuring}}
\label{subsec:existing algorithms}

We temporarily assume that all the random variables $\{\widetilde{h}_0,\widetilde{f}_n,\widetilde{g}_n\}$ are fixed as in the previous works \cite{Arun2020RFocus,ren2022configuring}. Thus, all the channels $\{h_0,h_n\}$ are complex constants, but their values are still unknown. The expectation operation $\mathbb E$ in problem \eqref{opt problem} can then be dropped. Let us go over how this deterministic version of problem \eqref{opt problem} is addressed in \cite{ren2022configuring}.

We use $\Delta_n$ to denote the phase difference between the direct channel $h_0$ and the reflected channel $h_n$:
\begin{equation}
\Delta_n=\angle h_0-\angle h_n,\quad\text{for}\; n=1,\ldots,N.
\end{equation}
Clearly, for the continuous beamforming case with $K\rightarrow\infty$, it is optimal to align each reflected channel with the direct channel in order to maximize the overall channel strength, so the optimal choice of $\theta_n$ is
\begin{equation}
    \theta^\star_n = \Delta_n.
\end{equation}
However, when $K<\infty$, the above solution may not be contained in the discrete set $\Phi_K$. A simple idea is to rotate every $h_n$ to the closest position to $h_0$ in the complex plane, namely the \emph{closest point projection (CPP)} method:
\begin{equation}
\theta_n^{\mathrm{CPP}}=\arg \min _{\theta_n \in \Phi_K}\left|\mathrm{Arg}\left(\frac{h_ne^{j\theta_n}}{h_0}\right)\right|,
\label{CPP}
\end{equation}
where $\mathrm{Arg}(\cdot)$ is the principal argument of a complex number. We can further find an approximation bound for the CPP algorithm as stated in the following lemma.
\begin{lemma}
\label{prop:approximation CPP without fading}
Suppose that the fading components $\{\widetilde{h}_0,\widetilde{f}_n,\widetilde{g}_n\}$ are all fixed in problem \eqref{opt problem} so that the expectation operation $\mathbb E$ can be dropped in the optimization objective. Let $f^\star$ be the global optimum of the resulting deterministic version of problem \eqref{opt problem}. If $h_0$ can be bounded away from zero, then the solution of the CPP algorithm in \eqref{CPP} satisfies
\begin{equation}
\label{CPP:bounds}
\cos^2(\frac{\pi}{K})\cdot f^\star\le\Bigg|h_0+\sum_{n=1}^{N}h_ne^{j\theta_n^{\mathrm{CPP}}}\Bigg|^2\le f^\star.
\end{equation}
\end{lemma}
Lemma \ref{prop:approximation CPP without fading} has been established in \cite[Proposition 2]{zhang2022configuring}. Actually, when channels are known, the discrete optimization in \eqref{opt problem} can be optimally solved in linear time in terms of $N$, as shown in \cite{ren2022linear}; but this linear-time optimal algorithm is irrelevant to blind beamforming, so we omit it in this paper.

\begin{example}
\label{example:CPP_K2}
    When $K=2$, i.e., when each $\theta_n\in \{0, \pi\}$, the lower bound equals zero in the worst-case scenario, as shown in the following example. Assume that all the magnitudes of all components $|h_n|$ are equal; assume also that $\Delta_n=\pi/2-\varepsilon$ for $n=1,\ldots,N/2$ and $\Delta_n=-\pi/2+\varepsilon$ for $n=N/2+1,\ldots,N$, where $\Delta_n =\angle h_0-\angle h_n$. Let $\varepsilon\rightarrow0^+$. According to \eqref{CPP}, we have $\theta_n^{\mathrm{CPP}}=0$ for every $n=1,\ldots,N$ and thus the two groups of reflected channels cancel out. Actually, the optimal solution in this case should be $\theta_n=0$ for $n=1,\ldots,N/2$ and $\theta_n=\pi$ for $n=N/2+1,\ldots,N$.
    But we emphasize that the above example is a crafted case. In most random realizations of the $K=2$ case, CPP still performs quite well, as shown in Section \ref{sec:Numerical}.
\end{example}
%\begin{IEEEproof}
%    The upper bound is evident. We just concentrate on the lower bound in what follows.
%    Denote by $\beta_n,n=0,1,\ldots,N$ and $\alpha_n,n=0,1,\ldots,N$ the strength and phase of $h_n$ respectively. Clearly we have $f^\star \leq (\sum_{n=0}^{N}\beta_n)^2$. We also have
%    \begin{align}
%    \bigg|h_0+\sum_{n=1}^{N}h_ne^{j\theta_n^{\mathrm{CPP}}}\bigg|^2 & =\left|\beta_0 e^{j \alpha_0}+\sum_{n=1}^N \beta_n e^{j\left(\theta_n^{\mathrm{CPP}}+\alpha_n\right)}\right|^2 \notag\\
%    & =\left|\beta_0+\sum_{n=1}^N \beta_n e^{j\left(\theta_n^{\mathrm{CPP}}-\Delta_n\right)}\right|^2 \notag\\
%    & \geq \left|\beta_0+\sum_{n=1}^N \beta_n \cos \left(\theta_n^{\mathrm{CPP}}-\Delta_n\right)\right|^2 \notag \\
%    & \stackrel{(a)}{\geq} \left|\beta_0+\sum_{n=1}^N \beta_n \cos \frac{\omega}{2}\right|^2 \notag\\
%    & \geq \cos ^2(\omega / 2) \left(\sum_{n=0}^N \beta_n
%    \right)^2 \notag\\
%    & \geq \cos^2(\frac{\pi}{K})\cdot f^\star,
%    \end{align}
%    where $(a)$ follows since $|\theta_n^{\mathrm{CPP}}-\Delta_n|\leq\omega/2$. The proof is then completed.
%\end{IEEEproof}

Notice that the CPP algorithm requires the information of $\{\Delta_n\}$ and thus requires channel estimation. The main result of \cite{Arun2020RFocus,ren2022configuring} is that CPP can be performed implicitly in the absence of channel information. We focus on describing the CSM algorithm in \cite{ren2022configuring} in what follows, since the RFocus algorithm \cite{Arun2020RFocus} has similar steps. To start, the CSM algorithm tries out $T$ random samples of $\bm \theta=(\theta_1,\ldots,\theta_N)$, the $t$th random sample denoted by $\bm \theta_t=(\theta_{1t},\ldots,\theta_{Nt})$. For the $t$th random sample, we measure the corresponding received signal power denoted by $|Y_t|^2$.
Next, the $T$ random samples are grouped with respect to each $n=1,\ldots,N$ and each $k=1,\ldots,K$:
\begin{equation}
    \mathcal{Q}_{n k} =\left\{t \mid \theta_{nt}=k\omega\right\},
\end{equation}
i.e., $\mathcal{Q}_{n k}$ is a set of indices of random samples in which the phase shift of the $n$th RE equals $k\omega$. Note that the same random sample appears in multiple $\mathcal Q_{nk}$'s with different $n$. After obtaining all the $\mathcal Q_{nk}$'s, we compute the empirical conditional average of the received signal power as
\begin{equation}
\widehat{\mathbb{E}}\left[|Y|^2 \mid \theta_n=k \omega\right]=\frac{1}{\left|\mathcal{Q}_{n k}\right|} \sum_{t \in \mathcal{Q}_{n k}}|Y_t|^2.
\label{eq:conditional_sample_mean}
\end{equation}
Intuitively, $\widehat{\mathbb{E}}\left[|Y|^2 \mid \theta_n=k \omega\right]$ characterizes the average performance of letting $\theta_n=k\omega$ while the rest phase shifts are chosen at random. We then decide each $\theta_n$ according to the average performance:
\begin{equation}
\theta_n^{\mathrm{CSM}}=\arg \max _{\varphi \in \Phi_K} \widehat{\mathbb{E}}\left[|Y|^2 \mid \theta_n=\varphi\right].
\label{eq:theta_selection_CSM}
\end{equation}
We illustrate the above steps with the following toy example.

\begin{example}
\label{exmple_of_CSM}
    Assume that $N=4$, $K=2$, and $T=6$. All the random samples and their corresponding received signal powers $|Y|^2$ are listed in Table~\ref{table example CSM}. We then compute the empirical conditional averages with respect to $\theta_1=0$ and with respect to $\theta_1=\pi$ as
    \begin{align}
        &\widehat{\mathbb E}\left[|Y|^2\mid\theta_{1}=0\right]=\frac{2.8+1.0+1.4}{3}=1.4\notag\\
        &\widehat{\mathbb E}\left[|Y|^2\mid\theta_{1}=\pi\right]=\frac{1.5+3.3+0.3}{3}=1.7.\notag
    \end{align}
    Thus, the CSM algorithm would let $\theta_1=\pi$ according to \eqref{eq:theta_selection_CSM}. The rest phase shifts can be decided similarly. The complete solution in this example is $\theta_1=\pi$, $\theta_2=0$, $\theta_3=\pi$, and $\theta_4=0$.
\begin{table}[t]
\renewcommand{\arraystretch}{1.2}
\small
\centering
   \caption{Toy example of CSM when $N=4$, $K=2$, and $T=6$.}
\label{table example CSM}
%\resizebox{!}{!}{
\begin{tabular}{|c|c|c|}
\hline
$t$ & $(\theta_1,\theta_2,\theta_3,\theta_4)$     & $|Y|^2$ \\ \hline
\hline
1     & $(0,\pi,0,0)$     & 2.8          \\ \hline
2     & $(0,0,0,0)$       & 1.0          \\ \hline
3     & $(\pi,\pi,\pi,0)$ & 1.5      \\ \hline
4     & $(\pi,0,\pi,\pi)$ & 3.3      \\ \hline
5     & $(\pi,\pi,0,\pi)$ & 0.3      \\ \hline
6     & $(0,0,\pi,\pi)$   & 0.4         \\ \hline
\end{tabular}
\end{table}
\end{example}

We now establish the equivalence between the CPP solution \eqref{CPP} and the CSM solution \eqref{eq:theta_selection_CSM}. As $T\rightarrow\infty$, the empirical conditional average converges to the actual conditional expectation, i.e., the value of $\widehat{\mathbb{E}}\left[|Y|^2 \mid \theta_n=k \omega\right]$ converges to
\begin{multline}
\label{conditional expectation}
    \mathbb{E}\left[|Y|^2 \mid \theta_n=k \omega\right]=2P|h_0||h_n|\cos(k\omega-\Delta_n)\\+P\sum^N_{m=1,m\neq n}|h_m|^2+\sigma^2.
\end{multline}
Evidently, in order to maximize $\mathbb{E}\left[|Y|^2 \mid \theta_n=k \omega\right]$, we need to minimize the gap between $k\omega$ and $\Delta_n$. In other words, we need to choose $\theta_n$ to be the point in $\{\omega,2\omega,\ldots,K\omega\}$ that is the closest to $\Delta_n$, namely the CPP algorithm in \eqref{CPP}. The above results can be immediately carried over to the RFocus algorithm \cite{Arun2020RFocus} by considering random samples of the {\footnotesize{ON-OFF}} statuses of the passive antennas.

\begin{remark}
    \label{remark:CSM_T}
    A more in-depth analysis of the CSM algorithm is provided in \cite{ren2022configuring}. It shows that the CPP solution equals the CSM solution with high probability so long as $T=\Omega(N^2(\log N)^3)$. Furthermore, if the average reflected signal power per RE is fixed, then an SNR boost of $\Theta(N^2)$ can be achieved by the CSM algorithm.
\end{remark}

\begin{remark}
   A common heuristic is to simply pick the best random sample we have tested so far from the codebook, which amounts to the beam training method \cite{you2020fast,ning2021terahertz,wang2022fast,wang2021jointbeam} with the codewords uniformly generated in the codebook. Nevertheless,   
   \cite{ren2022configuring} shows that the SNR boost by this method is only $\Theta(N\log T)$. Thus, the CSM algorithm outperforms the beam training method in general.
\end{remark}

\begin{remark}[Preview of Why RFocus and CSM may Fail]
\label{remark:failure}
    In the above analysis, we have implicitly assumed that the direct channel strength $|h_0|$ is a strictly positive number. Now let us consider the extreme case in which $|h_0|=0$. The conditional expectation of the received signal power in \eqref{conditional expectation} reduces to
    \begin{equation}
    \mathbb{E}\left[|Y|^2 \mid \theta_n=k \omega\right]=P\sum^N_{m=1,m\neq n}|h_m|^2+\sigma^2.
    \end{equation}
    Observe that the value of $\mathbb{E}\left[|Y|^2 \mid \theta_n=k \omega\right]$ now becomes a constant regardless of the choice of $\theta_n$. Thus, we can no longer decide the optimal choice of $\theta_n$ by comparing the empirical conditional averages as in \eqref{eq:theta_selection_CSM}, and consequently the CSM algorithm fails to work. A formal analysis of this issue is presented in Section \ref{subsec failure}.
\end{remark}

%The equivalence between CSM and CPP can be established as follows. When the number of random samples $T\to\infty$, the conditional sample mean in \eqref{eq:conditional_sample_mean} is tend to the corresponding conditional expectation, which is given by $\mathbb{E}\left[|Y|^2 \mid \theta_n=k \omega\right]=2P|h_0||h_n|\cos(k\omega-\Delta_n)+P\sum_{m\neq n}|h_n|+\sigma^2$. Clearly, when treated as a function of $k\omega$, the above conditional expectation is maximized when the gap between $k\omega$ and $\Delta_n$ modulo $2\pi$ is minimized; this optimal $k\omega$ can be recognized as the CPP solution in \eqref{CPP}.

%It has been shown in \cite{ren2022configuring} that the CSM algorithm can yield a quadratic overall channel power boost, i.e., $\Theta(N^2)$ boost, when the number of random samples satisfies $T=\Omega(n^2(\log N)^3)$, while the random max sampling (RMS) method can only yields $\Theta(N\log T)$ boost when $T=o(\sqrt{N})$.

\subsection{Extension to Fading Channels}
\label{subsec:CSM fading}

The previous works \cite{Arun2020RFocus,ren2022configuring} verify the performance of their blind beamforming algorithms in the context of static channels, i.e., when the channels $\{h_0,h_n\}$ are all fixed. Our new result here is to show that the RFocus algorithm \cite{Arun2020RFocus} and the CSM algorithm \cite{ren2022configuring} continue to work for the fading channels in \eqref{Rician channel model} so long as the direct channel is LoS, i.e., when $\mathbb E[h_0]$ is sufficiently large to be bounded away from zero.

For the fading channel model in \eqref{Rician channel model}, we define 
\begin{equation}
\Delta_n=\angle \overline{h}_0-\angle \overline{h}_n,\quad\text{for}\; n=1,\ldots,N.
\end{equation}
The conditional expectation of the received signal power now becomes
\begin{align}
\label{EY^2}
&\mathbb{E}\left[|Y|^2\mid\theta_n=\varphi\right]\notag\\
%&=P\mathbb{E}\left[\bigg|h_0+h_ne^{j\varphi}+\sum_{m\neq n}h_me^{j\theta_m}\bigg|^2\right]+\sigma^2 \notag \\
&=2P\sqrt{\frac{\gamma_{00}\gamma_{0n}\gamma_{n0}\delta_{00}\delta_{0n}\delta_{n0}}{(1+\delta_{00})(1+\delta_{0n})(1+\delta_{n0})}}\cos(\varphi-\Delta_n)+\sigma^2\,+\notag\\
&\quad\;P\left(\gamma_{00}+\sum_{n=1}^{N}\frac{\gamma_{0n}\gamma_{n0}(\delta_{0n}\delta_{n0}+\delta_{0n}+\delta_{n0}+1)}{(1+\delta_{0n})(1+\delta_{n0})} \right),
\end{align}
where the expectation is taken over all random fading components and the randomly selected $\theta_m,m\neq n$. Because of the LoS assumption, we have $\mathbb E[h_0]$ be sufficiently large and thereby prevent the coefficient $\sqrt{\frac{\gamma_{00}\gamma_{0n}\gamma_{n0}\delta_{00}\delta_{0n}\delta_{n0}}{(1+\delta_{00})(1+\delta_{0n})(1+\delta_{n0})}}$ from tending to zero. Following the argument in Section \ref{subsec:existing algorithms}, we treat $\mathbb{E}\left[|Y|^2\mid\theta_n=\varphi\right]$ as a function of $\varphi$, and then find that maximizing its value amounts to minimizing the gap between $\varphi$ and $\Delta_n$. Moreover, notice that $\widehat{\mathbb{E}}\left[|Y|^2 \mid \theta_n=\varphi\right]$ converges to $\mathbb{E}\left[|Y|^2\mid\theta_n=\varphi\right]$ as $T\rightarrow\infty$. Thus, the solution by the CSM algorithm is equivalent to the solution by the CPP algorithm:
\begin{equation}
\label{fading CPP}
    \theta_n^{\mathrm{CPP}}=\arg \min _{\theta_n \in \Phi_K}\left|\mathrm{Arg}\left(\frac{\overline{h}_ne^{j\theta_n}}{\overline{h}_0}\right)\right|,
\end{equation}
when $T\to \infty$. The following proposition formalizes the above result.

\begin{proposition}
\label{prop:equivalent_fading}
For $K\geq 3$, for any fixed $\xi\in(0,1)$, the CSM solution $\bm \theta^{\mathrm{CSM}}$ in \eqref{eq:theta_selection_CSM} equals the CPP solution $\bm \theta^{\mathrm{CPP}}$ in \eqref{fading CPP} with a probability of at least $1-\xi$ so long as $T=\Omega(N^3\log N)$.
\end{proposition}
\begin{IEEEproof}
Note that we would have $\theta^{\mathrm{CSM}}_n=\theta^{\mathrm{CPP}}_n$ when $\widehat{\mathbb{E}}\left[|Y|^2 \mid \theta_n=k\omega\right]$ and $\mathbb{E}\left[|Y|^2 \mid \theta_n=k\omega\right]$ are sufficiently close to each other for all $k$. Let $\chi_{n 1}$ and $\chi_{n 2}$ be the largest and the second-largest values of $\cos \left(k \omega-\Delta_n\right)$, respectively, for each $k=1, \ldots, K$; the difference between $\chi_{n 1}$ and $\chi_{n 2}$ is denoted by $\epsilon_n=\chi_{n 1}-\chi_{n 2}$. The above observation can be formulated as:
\begin{equation}
\label{eq:condition}
\left|\widehat{\mathbb{E}}\left[|Y|^2 \mid \theta_n=k \omega\right]-\mathbb{E}\left[|Y|^2 \mid \theta_n=k \omega\right]\right|<2 \beta P \epsilon_n
\end{equation}
holds for every $k$, where
\begin{equation}
\beta=\sqrt{\frac{\gamma_{00}\gamma_{0n}\gamma_{n0}\delta_{00}\delta_{0n}\delta_{n0}}{(1+\delta_{00})(1+\delta_{0n})(1+\delta_{n0})}}.
\end{equation}
In the remainder of the proof, we discuss in what regime of $(N, T)$ the condition \eqref{eq:condition} holds for all $n=1, \ldots, N$.

Without loss of generality, we focus on a particular $(n, k)$ and its corresponding conditional subset $\mathcal{Q}_{n k}$. Let $T_{n k}$ be the cardinality of $\mathcal{Q}_{n k}$. With each $\theta_{n t}$ drawn from $\Phi_K$ uniformly and independently, we have
\begin{equation}
T_{n k}=\frac{T}{K} \quad \text { with high probability. }
\end{equation}
Notice that the received signal $Y$ can be formulated as
\begin{equation}
    Y = \overline{Y} + \widetilde
    {Y} + Z,
\end{equation}
where
\begin{subequations}
\begin{align}
&\overline{Y}=\left(\sqrt{\frac{\gamma_{00}\delta_{00}}{1+\delta_{00}}}\overline{h}_0+\sum_{n=1}^N\sqrt{\frac{\gamma_{0n}\gamma_{n0}\delta_{0n}\delta_{n0}}{(1+\delta_{0n})(1+\delta_{n0})}}\overline{h}_ne^{j\theta_n}\right)X \notag\\
&\widetilde{Y}=\left(\sqrt{\frac{\gamma_{00}}{1+\delta_{00}}}\widetilde{h}_0+\sum_{n=1}^N\sqrt{\frac{\gamma_{0n}\gamma_{n0}}{(1+\delta_{0n})(1+\delta_{n0})}}\widetilde{h}_ne^{j\theta_n}\right)X,\notag
\end{align}
\end{subequations}
are the parts of the received signal related to the LoS components and the NLoS components, respectively.

To ease notation in the further discussion, we define a sequence of new variables:
\begin{align}
    \eta_{nk} &= \mathbb{E}\left[\left|\overline{Y}\right|^2\mid\theta_n=k\omega\right]\\
    \hat{\eta}_{nk} &=\frac{1}{T_{nk}}\sum_{t\in\mathcal{Q}_{nk}}\left|\overline{Y}_t\right|^2\\
     \sigma^2_{nk} &=\mathbb{E}\left[\left|\widetilde{Y} +Z\right|^2\mid\theta_n=k\omega\right]\\
     \hat{\sigma}^2_{nk} &= \frac{1}{T_{nk}}\sum_{t\in\mathcal{Q}_{nk}}\left|\widetilde{Y}_t+Z_t\right|^2 \\
     \delta_{nk} &= \frac{2}{T_{nk}}\sum_{t\in\mathcal{Q}_{nk}}\mathfrak{Re}\left\{\overline{Y}_t^H (\widetilde{Y}_t+Z_t)\right\}.
\end{align}
It can be shown that
\begin{subequations}
\begin{align}
\mathbb{E}\left[|Y|^2 \mid \theta_n=k \omega\right] & =\eta_{n k}+\sigma_{n k}^2 \\
\widehat{\mathbb{E}}\left[|Y|^2 \mid \theta_n=k \omega\right] & =\hat{\eta}_{n k}+\hat{\sigma}_{n k}^2+\delta_{n k}.
\end{align}
\end{subequations}
Now, for each pair $(n, k)$, we aim to bound the probability of the error event
\begin{equation}
\mathcal{E}_{n k}=\left\{\left|\widehat{\mathbb{E}}\left[|Y|^2 \mid \theta_n=k \omega\right]-\mathbb{E}\left[|Y|^2 \mid \theta_n=k \omega\right]\right| \geq \epsilon_0\right\} \notag
\end{equation}
where the constant $\epsilon_0=\min _{n=1, \ldots, N}\left\{2 \beta\epsilon_n\right\}$. We show that
\begin{align}
 \mathbb{P}\left\{\mathcal{E}_{n k}\right\} 
& = \mathbb{P}\left\{\left|\hat{\eta}_{n k}+\hat{\sigma}_{n k}^2+\delta_{n k}-\eta_{n k}-\sigma_{nk}^2\right| \geq \epsilon_0\right\} \notag\\
& \stackrel{(a)}{\leq} \mathbb{P}\left\{\left|\hat{\eta}_{n k}-\eta_{n k}\right| \geq \frac{\epsilon_0}{3}\right\}+\mathbb{P}\left\{\left|\hat{\sigma}_{n k}^2-\sigma_{nk}^2\right| \geq \frac{\epsilon_0}{3}\right\} \notag\\
& \qquad+\mathbb{P}\left\{\left|\delta_{n k}\right| \geq \frac{\epsilon_0}{3}\right\},
\label{Epsilon bound}
\end{align}
where $(a)$ follows by the fact that $|a_1+a_2+a_3|\geq\epsilon_0$ implies that at least one $|a_i|\geq\epsilon_0/3$.

We now wish to quantify the upper bound in \eqref{Epsilon bound} in closed form. Toward this end, 
let us consider the following four error events parameterized by $q>0$ and $\epsilon>0$:
\begin{subequations}
\begin{align}
& \mathcal{E}_{n k, 1}(q)=\left\{\eta_{n k} \geq qP\nu\right\} \\
& \mathcal{E}_{n k, 2}(\epsilon)=\left\{\left|\hat{\eta}_{n k}-\eta_{n k}\right| \geq \epsilon\right\}\\
& \mathcal{E}_{n k, 3}(\epsilon)=\left\{\left|\hat{\sigma}_{n k}^2-\sigma_{nk}^2\right| \geq \epsilon\right\}\\
& \mathcal{E}_{n k, 4}(\epsilon)=\left\{\left|\delta_{n k}\right| \geq \epsilon\right\},
\end{align}
\end{subequations}
where
\begin{equation}
    \nu=\frac{\gamma_{00}\delta_{00}}{1+\delta_{00}}+\sum_{n=1}^N\frac{\gamma_{0n}\gamma_{n0}\delta_{0n}\delta_{n0}}{(1+\delta_{0n})(1+\delta_{n0})}.
\end{equation}
The probabilities of the above error events can be upper bounded as
\begin{subequations}
\begin{align}
    \mathbb{P}\left\{\mathcal{E}_{n k, 1}(q)\right\} &\stackrel{(a)}{\leq} 4 e^{-q / 4}, \label{eq:pe1}\\
    \mathbb{P}\left\{\mathcal{E}_{n k, 2}(\epsilon) \mid \mathcal{E}_{n k, 1}^c(q)\right\} &\stackrel{(b)}{\leq} 2 \exp \left(-\frac{2 \epsilon^2  T_{n k}}{q^2 P^2\nu^2}\right), \label{eq:pe2}\\
    \mathbb{P}\left\{\mathcal{E}_{n k, 3}(\epsilon)\right\} &\stackrel{(c)}{\leq} \frac{\widetilde{\sigma}^2}{\epsilon^2 T_{n k}}, \label{eq:pe3}\\
    \mathbb{P}\left\{\mathcal{E}_{n k, 4}(\epsilon) \mid \mathcal{E}_{n k, 1}^c(q)\right\} &\stackrel{(d)}{\leq} \frac{2 q P \nu \widetilde{\sigma}}{\epsilon^2 T_{n k}},\label{eq:pe4}
\end{align}
\end{subequations}
where
\begin{equation}
    \widetilde{\sigma} = P\left(\sum_{n=1}^N\frac{\gamma_{0n}\gamma_{n0}(\delta_{0n}+\delta_{n0}+1)}{(1+\delta_{0n})(1+\delta_{n0})}+\frac{\gamma_{00}}{1+\delta_{00}}\right)+\sigma^2.
\end{equation}
In the above bounds, $(a)$ follows by \cite[Lemma 2]{ren2022configuring}, $(b)$ follows by Hoeffding's inequality, and $(c)$ and $(d)$ follow by Chebyshev's inequality.

Putting \eqref{eq:pe1}--\eqref{eq:pe4} together, we then arrive at a closed-form upper bound on the original error probability as
\begin{align}
& \mathbb{P}\left\{\mathcal{E}_{n k}\right\} \notag\\
&\leq \mathbb{P}\left\{\mathcal{E}_{n k, 2}\left(\frac{\epsilon_0} 3\right)\right\}+\mathbb{P}\left\{\mathcal{E}_{n k, 3}\left(\frac{\epsilon_0} 3\right)\right\}+\mathbb{P}\left\{\mathcal{E}_{n k, 4}\left(\frac{\epsilon_0} 3\right)\right\} \notag\\
& \leq \mathbb{P}\left\{\mathcal{E}_{n k, 2}\left(\frac{\epsilon_0} 3\right) \Big|\, \mathcal{E}_{n k, 1}^c(q)\right\}+\mathbb{P}\left\{\mathcal{E}_{nk, 3}\left(\frac{\epsilon_0} 3\right) \Big|\,\mathcal{E}_{n k, 1}^c(q)\right\} \notag\\
&\quad+\mathbb{P}\left\{\mathcal{E}_{n k, 4}\left(\frac{\epsilon_0} 3\right)\right\}+2 \times\mathbb{P}\left\{\mathcal{E}_{n k, 1}(q)\right\} \notag\\
&\leq 2 \exp \left(-\frac{2 \epsilon_0^2  T}{9 q^2 P^2 \nu^2K}\right)+\frac{9 \widetilde{\sigma}^2 K}{\epsilon_0^2 T}+\frac{18 q P\nu \widetilde{\sigma}K}{ \epsilon_0^2 T} +8e^{-q/4}.
\end{align}

To further obtain the regime of $(N, T)$ within which the condition \eqref{eq:condition} holds for all $n=1, \ldots, N$, we now consider the overall error event $\mathcal{E}_0=$ $\bigcup_{(n, k)} \mathcal{E}_{n k}$, the probability of which can be bounded by the union bound as
\begin{equation}
\begin{aligned}
    \mathbb{P}\left\{\mathcal{E}_0\right\}
    &\leq \sum_{n=1}^N\sum_{k=1}^K\mathbb{P}\{\mathcal{E}_{n k}\}\\
    & \leq 2NK \exp \left(-\frac{2 \epsilon_0^2  T}{9 q^2 P^2 \nu^2K}\right)+\frac{9 \widetilde{\sigma}^2 NK^2}{\epsilon_0^2 T}\notag\\
    &\qquad\qquad+\frac{18 q P\nu \widetilde{\sigma}NK^2}{ \epsilon_0^2 T} +8 N K e^{-q / 4}.
\end{aligned}
\end{equation}
In the meanwhile, for any fixed $0<\xi<1$ and any  $0<p_0<\xi/4$, we have
\allowdisplaybreaks
\begin{subequations}
\begin{align}
2 N K \exp \left(-\frac{2 \epsilon_0^2  T}{9 q^2 P^2 \nu^2K}\right) & \leq p_0 \text { if } T=\Omega\left(N^2 q^2 \log N\right), \\
\frac{9 \widetilde{\sigma}^2 NK^2}{\epsilon_0^2 T} & \leq p_0 \text { if } T=\Omega(N^3), \\
\frac{18 q P\nu \widetilde{\sigma}NK^2}{ \epsilon_0^2 T} & \leq p_0 \text { if } T=\Omega\left(N^3 q\right), \\
8 N K e^{-q / 4} & \leq p_0 \text { if } q=\Omega(\log N).
\end{align}
\end{subequations}
Because $N^3\log N>N^2(\log N)^3$, after combining the above result, we obtain that $\mathbb{P}\left\{\mathcal{E}_0\right\}<\xi$ whenever $T=\Omega(N^3\log N)$. 
Consequently, for the fading channel case, we have $\mathbb{P}\left\{\bm\theta^{\mathrm{CSM}}=\bm\theta^{\mathrm{CPP}}\right\}\geq1-\xi$ whenever $T=\Omega(N^3\log N)$. The proof is then completed.
\end{IEEEproof}

With the above result, we then extend the result of Lemma \ref{prop:approximation CPP without fading} to the CSM algorithm for the fading channel case as follows:
\begin{proposition}
\label{prop:CSM fading}
If $\mathbb E[h_0]$ can be bounded away from zero, then the solution of the CSM algorithm in \eqref{eq:theta_selection_CSM} satisfies
\begin{equation}
\label{CPP:bounds_fading}
(1-\xi)\cos^2(\frac{\pi}{K})\cdot f^\star\le\mathbb E\bigg[\bigg|h_0+\sum_{n=1}^{N}h_ne^{j\theta_n^{\mathrm{CSM}}}\bigg|^2\bigg]\le f^\star
\end{equation}
for any fixed $\xi\in(0,1)$
so long as $T=\Omega(N^3\log N)$, where $f^\star$ represents the global optimum of problem \eqref{opt problem}.
\end{proposition}
\begin{IEEEproof}
The upper bound is evident. We focus on establishing the lower bound. First, we have
\allowdisplaybreaks
\begin{multline}
f^\star\leq\left(\sqrt{\frac{\gamma_{00}\delta_{00}}{1+\delta_{00}}}+\sum_{n=1}^{N}\sqrt{\frac{\gamma_{0n}\gamma_{n0}\delta_{0n}\delta_{n0}}{(1+\delta_{0n})(1+\delta_{n0})}}\right)^2 \\
    +\frac{\gamma_{00}}{1+\delta_{00}}+\sum_{n=1}^{N}\frac{\gamma_{0n}\gamma_{n0}(\delta_{0n}+\delta_{n0}+1)}{(1+\delta_{0n})(1+\delta_{n0})}
\end{multline}
by assuming that every $\overline{h}_n$ has been aligned perfectly with $\overline{h}_0$.
It can be further shown that the CPP solution satisfies

\begin{align}  &\mathbb{E}\left[\bigg|h_0+\sum_{n=1}^{N}h_ne^{j\theta_n^{\mathrm{CPP}}}\bigg|^2\right] \notag\\
    &=\left|\sqrt{\frac{\gamma_{00}\delta_{00}}{1+\delta_{00}}}+\sum_{n=1}^{N}\sqrt{\frac{\gamma_{0n}\gamma_{n0}\delta_{0n}\delta_{n0}}{(1+\delta_{0n})(1+\delta_{n0})}}e^{j(\theta_n^{\mathrm{CPP}}-\Delta_n)}\right|^2 \notag\\
    &\quad+\frac{\gamma_{00}}{1+\delta_{00}}+\sum_{n=1}^{N}\frac{\gamma_{0n}\gamma_{n0}(\delta_{0n}+\delta_{n0}+1)}{(1+\delta_{0n})(1+\delta_{n0})} \notag \\
    &\geq \cos^2(\frac{\pi}{K})\cdot\left(\sqrt{\frac{\gamma_{00}\delta_{00}}{1+\delta_{00}}}+\sum_{n=1}^{N}\sqrt{\frac{\gamma_{0n}\gamma_{n0}\delta_{0n}\delta_{n0}}{(1+\delta_{0n})(1+\delta_{n0})}}\right)^2\notag \\
    &\quad+\cos^2(\frac{\pi}{K})\cdot\left(\frac{\gamma_{00}}{1+\delta_{00}}+\sum_{n=1}^{N}\frac{\gamma_{0n}\gamma_{n0}(\delta_{0n}+\delta_{n0}+1)}{(1+\delta_{0n})(1+\delta_{n0})}\right) \notag \\
    &\geq \cos^2(\frac{\pi}{K})\cdot f^\star.
    \label{cos:bound}
\end{align}
Finally, we show that
\begin{align}
&\mathbb{E}\left[\bigg|h_0+\sum_{n=1}^{N}h_ne^{j\theta_n^{\mathrm{CSM}}}\bigg|^2\right]\notag\\
&\ge
\mathbb P\Big\{\bm\theta^{\text{CSM}}=\bm\theta^{\text{CPP}}\Big\}\times\mathbb{E}\left[\bigg|h_0+\sum_{n=1}^{N}h_ne^{j\theta_n^{\mathrm{CPP}}}\bigg|^2\right]\notag\\
&\overset{(a)}{\ge}
(1-\xi)\cdot\mathbb{E}\left[\bigg|h_0+\sum_{n=1}^{N}h_ne^{j\theta_n^{\mathrm{CPP}}}\bigg|^2\right]\notag\\
&\overset{(b)}{\ge} (1-\xi)\cos^2(\frac{\pi}{K})\cdot f^\star.
\end{align}
where $(a)$ follows by Proposition \ref{prop:equivalent_fading}, $(b)$ follows by the inequality in \eqref{cos:bound}.
\end{IEEEproof}

The above result can be readily extended for the RFocus algorithm. We remark that the LoS assumption is critical to the above conclusion, without which the existing beamforming algorithms may fail to work, as discussed in the sequel. %The following proposition summarizes the above discussion.

%Clearly, when treated as a function of $\varphi$, the above conditional expectation is maximized when the gap between $\varphi$ and $\Delta_n$ modulo $2\pi$ is minimized; this optimal $\varphi$ can also be recognized as the CPP solution in \eqref{CPP} when $\Delta_n$ is now defined as the phase different between each $h_n$ and $h_0$ in terms of their fixed fading components, i.e., $\Delta_n=\angle\overline{h}_0-\angle\overline{h}_n,\;\text{for}\; n=1,\ldots,N.$

%Thus, the blind beamforming algorithms in \cite{Arun2020RFocus,ren2022configuring} can still work in the LoS fading channel case. Moreover, it has been shown in \cite{ren2022configuring} that the above algorithm can yield a quadratic overall channel power boost, i.e., $\Theta(N^2)$ boost, when the number of random samples $T\to\infty$. The same conclusion also holds for the LoS fading channels, as shown in the following corollary.
%\begin{proposition}
%If $K>2$ and $\mathbb E[h_0]$ is bounded away from zero, then the RFocus algorithm \cite{Arun2020RFocus} and the CSM algorithm \cite{ren2022configuring} are equivalent to the CPP algorithm. The resulting expectation of the overall channel power in the long run achieved by CSM algorithm satisfies
%\begin{equation}	\mathbb{E}\left[\left|h_0+\sum_{n=1}^{N}h_ne^{j\theta_n^{\mathrm{CSM}}}\right|^2\right]=\Theta(N^2)
%\label{CSM_bound}
%\end{equation}
%as $T\rightarrow\infty$, where the expectation is taken over both random samples of $\boldsymbol{\theta}$ as well as random fading channels.
%\label{prop:CSM_order_bound}
%\end{proposition}

\subsection{Failure of RFocus \cite{Arun2020RFocus} and CSM \cite{ren2022configuring}}
\label{subsec failure}

We now point out a fundamental flaw in the existing blind beamforming algorithms. Let us begin with the extreme case where $\mathbb E[h_0]=0$. Notice that $\mathbb E[h_0]=0$ implies either $\gamma_{00}=0$ or $\delta_{00}=0$. Consequently, the conditional expectation in \eqref{EY^2} reduces to
\begin{align}
\mathbb{E}\left[|Y|^2\mid\theta_n=\varphi\right]
&= P\sum_{n=1}^{N}\frac{\gamma_{0n}\gamma_{n0}(\delta_{0n}\delta_{n0}+\delta_{0n}+\delta_{n0}+1)}{(1+\delta_{0n})(1+\delta_{n0})}\notag\\
&\qquad+\sigma^2+P\gamma_{00},
\end{align}
which does not depend on the choice of $\theta_n$. Thus, even if the $\widehat{\mathbb{E}}\left[|Y|^2 \mid \theta_n=k \omega\right]$ converges to $\mathbb{E}\left[|Y|^2 \mid \theta_n=k \omega\right]$, we cannot decide each $\theta_n$ according to the CSM operation in \eqref{eq:theta_selection_CSM}. But what if $\mathbb E[h_0]$ tends to zero? To see the answer, we now expand the empirical conditional average in \eqref{eq:conditional_sample_mean} as
\begin{equation}
    \widehat{\mathbb{E}}\left[|Y|^2 \mid \theta_n=\varphi\right]
    =2P\beta_1\cos(\varphi-\Delta_n)+\sum^7_{i=1}C_i,
\end{equation}
where
\allowdisplaybreaks
\begin{align*}
    &C_1 = \frac{2P}{\left|\mathcal{Q}_{n k}\right|}\sum_{t \in \mathcal{Q}_{n k}}\sum_{n=0}^{N}|h_{nt}|^2\\
    &C_2 = \frac{2P}{\left|\mathcal{Q}_{n k}\right|}\sum_{t \in \mathcal{Q}_{n k}}\mathfrak{Re}\left\{\beta_2\overline{h}^H_0\widetilde{h}_{nt}+\beta_3\widetilde{h}^H_{0t}\overline{h}_n+\beta_4\widetilde{h}^H_{0t}\widetilde{h}_{nt}\right\}\\
    &C_3 = \frac{2P}{\left|\mathcal{Q}_{n k}\right|}\sum_{t \in \mathcal{Q}_{n k}}\mathfrak{Re}\Bigg\{h^H_{0t}\sum_{m=1,m\neq n}^{N}h_{mt}e^{j\theta_{mt}}\Bigg\} \\
    &C_4 = \frac{2P}{\left|\mathcal{Q}_{n k}\right|}\sum_{t \in \mathcal{Q}_{n k}}\mathfrak{Re}\Bigg\{h^H_{nt}\sum_{m=1,m\neq n}^{N}h_{mt}e^{j(\theta_{mt}-\varphi)}\Bigg\}\\
    &C_5 = \frac{2 P}{\left|\mathcal{Q}_{n k}\right|}  \sum_{t \in \mathcal{Q}_{n k}} \mathfrak{Re}\Bigg\{\sum_{a=1}^N \sum_{b=1, b \neq a}^N h_{at} h^H_{bt} e^{j\left(\theta_{a t}-\theta_{b t}\right)}\Bigg\} \\
    &C_6 =\frac{2 \sqrt{P}}{\left|\mathcal{Q}_{n k}\right|} \sum_{t \in \mathcal{Q}_{n k}}\mathfrak{Re}\Bigg\{\sum_{n=0}^N h_{nt} e^{j \theta_{nt}} Z^H_{t}\Bigg\}\\
    &C_7 =\frac{1}{\left|\mathcal{Q}_{n k}\right|} \sum_{t \in \mathcal{Q}_{n k}}\left|Z_{t}\right|^2 \\
    & \beta_1=\sqrt{\frac{\gamma_{00}\gamma_{0n}\gamma_{n0}\delta_{00}\delta_{0n}\delta_{n0}}{(1+\delta_{00})(1+\delta_{0n})(1+\delta_{n0})}} \\
    &\beta_2 = \sqrt{\frac{\gamma_{00}\gamma_{0n}\gamma_{n0}\delta_{00}}{(1+\delta_{00})(1+\delta_{0n})(1+\delta_{n0})}} \\
    &\beta_3 = \sqrt{\frac{\gamma_{00}\gamma_{0n}\gamma_{n0}\delta_{0n}\delta_{n0}}{(1+\delta_{00})(1+\delta_{0n})(1+\delta_{n0})}} \\
    &\beta_4 = \sqrt{\frac{\gamma_{00}\gamma_{0n}\gamma_{n0}}{(1+\delta_{00})(1+\delta_{0n})(1+\delta_{n0})}}.
\end{align*}
Recall that the main idea of the CSM algorithm is to mimic the CPP solution. Toward this end, it requires that $\widehat{\mathbb{E}}\left[|Y|^2 \mid \theta_n=\theta_n^{\mathrm{CPP}}\right]>\widehat{\mathbb{E}}\left[|Y|^2 \mid \theta_n=\varphi\right]$ holds for any $\varphi\neq\theta_n^{\mathrm{CPP}}$ since we choose each $\theta_n$ to maximize the empirical conditional average as in \eqref{eq:theta_selection_CSM}; it is sufficient to require that
\begin{equation}
\left|\widehat{\mathbb{E}}\left[|Y|^2 \mid \theta_n=\varphi\right]-{\mathbb{E}}\left[|Y|^2 \mid \theta_n=\varphi\right]\right|<2 \beta_1 \epsilon_n,
\label{eq:suf condition}
\end{equation}
where $\epsilon_n>0$ is the difference between the highest value and the second highest value of $\cos \left(\varphi-\Delta_n\right)$ across all possible $\varphi \in \Phi_K$. Intuitively speaking, if each empirical conditional average is close to the actual conditional expectation, then we will not get confused with the actual $\theta_n$ that maximizes ${\mathbb{E}}\left[|Y|^2 \mid \theta_n=\varphi\right]$. 
Based on the above observation, we bound the error probability as
\begingroup
\allowdisplaybreaks
\begin{align}
    & \mathbb{P}\left\{\boldsymbol{\theta}^{\mathrm{CSM}} \neq \boldsymbol{\theta}^{\mathrm{CPP}}\right\} \notag \\
    & \stackrel{(a)}{\leq} \sum_{n=1}^N \mathbb{P}\left\{\theta_n^{\mathrm{CSM}} \neq \theta_n^{\mathrm{CPP}}\right\} \notag\\
    & \leq \sum_{n=1}^N \mathbb P\Big\{\Big|\widehat{\mathbb{E}}\left[|Y|^2 | \theta_n=\varphi\right]-{\mathbb{E}}\left[|Y|^2 |\theta_n=\varphi\right]\Big|>2 \beta_1 \epsilon_n\Big\} \notag\\
    & \stackrel{(b)}{\leq} \sum_{n=1}^N \frac{\operatorname{Var}(\sum^7_{i=1}C_i)}{4\left|\mathcal{Q}_{n k}\right| \beta_1^2 \epsilon_n^2} \notag \\
    & \stackrel{(c)}{\simeq}\sum_{n=1}^N \frac{K(1+\delta_{00})(1+\delta_{0n})(1+\delta_{n0})\operatorname{Var}(\sum^7_{i=1}C_i)}{4 T \gamma_{00}\gamma_{0n}\gamma_{n0}\delta_{00}\delta_{0n}\delta_{n0}\epsilon_n^2},
    \label{Pe_upper_bound}
\end{align}
\endgroup
where $(a)$ follows by the union bound, $(b)$ follows by Chebyshev's inequality, and $(c)$ follows since $\left|\mathcal{Q}_{n k}\right| \approx T / K$. 

The above upper bound suggests that just letting $T\rightarrow\infty$ is not enough for the CSM algorithm to work in the NLoS direct channel case. Rather, it requires
\begin{equation}
\label{Big_T}
    T = \Omega\left(\frac{1}{\gamma_{00}\delta_{00}}\right).
\end{equation}
In the next section, we propose an improved version of CSM algorithm called \emph{Grouped Conditional Sample Mean (GCSM)}, which (i) still works when $\mathbb E[h_0]=0$ and (ii) requires much fewer random samples than \eqref{Big_T} when $\mathbb E[h_0]\rightarrow0$.
\begin{remark}
The above result implies that we can tell the status (LoS or NLoS) of the direct channel without CSI. We just compare the values of $\widehat{\mathbb{E}}\left[|Y|^2 \mid \theta_n=\varphi\right]$ with different $\varphi$; the direct channel is NLoS if these empirical conditional averages are close to each other, and is LoS otherwise.
\end{remark}

\section{Adaptive Blind Beamforming by Grouping}
\label{sec:alg_NLoS}

This section introduces the main result of this paper, an adaptive blind beamforming algorithm that works for all channel cases regardless of the strength of the direct channel. Furthermore, we discuss how the proposed algorithm can be extended to multiple users.

\subsection{Grouped Conditional Sample Mean (GCSM)}
\begin{figure}
    \centering
    \includegraphics[width=9cm]{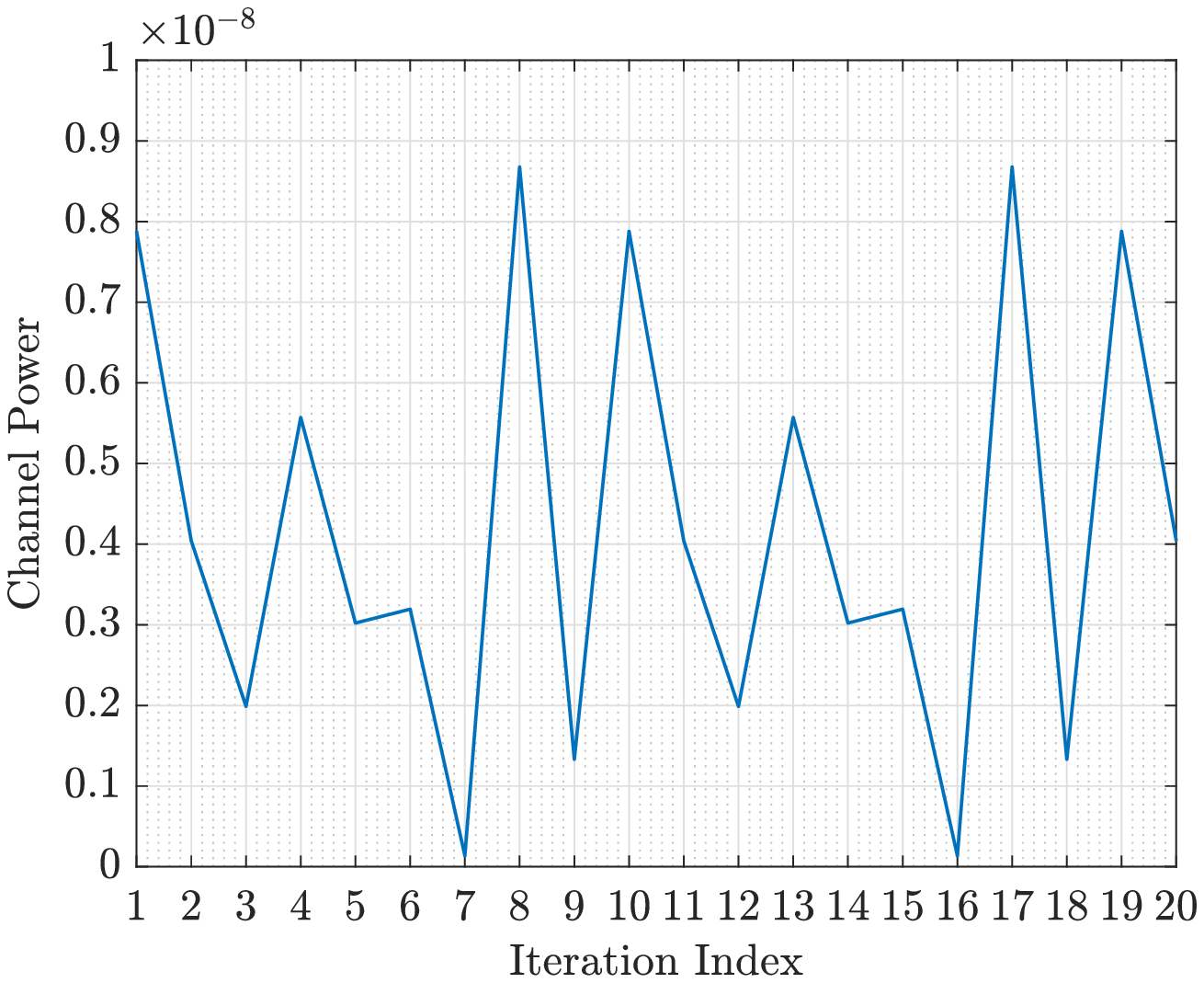}
    \caption{An example of the alternating CSM algorithm with the REs divided into two groups. Assume that $N=4$ and the channels are
   $[h_0,h_1,h_2,h_3,h_4]=[0,1.7646+2.1012j,0.2792-1.6644j,0.7178+3.1842j,0.6117-2.2282j]\times10^{-5}$; the reflected channels are grouped as $S_{\mathrm{I}}=\{h_1, h_4\}$ and $S^c_{\mathrm{I}}=\{h_2, h_3\}$. We assume that $T\rightarrow\infty$ and thus each empirical conditional average has converged to the actual conditional expectation.}
    \label{fig:alternative_CSM}
\end{figure}

Since $|\mathbb E[h_0]|$ being close to zero is the cause of the failure of the existing blind beamforming algorithms as stated in Section \ref{subsec failure}, a natural idea is to combine some reflected channels with $h_0$ so that the new virtual direct channel is sufficiently strong. Following this idea, a naive algorithm is to divide the REs into two groups and perform CSM algorithm between the two groups alternatingly. To be more specific, in each iteration, we optimize $\theta_n$'s for the REs in one group by CSM while holding the rest phase shifts fixed. But the performance of this alternating CSM algorithm cannot be justified, as illustrated by a counterexample in Fig.~\ref{fig:alternative_CSM}. Observe that the alternating CSM algorithm cannot guarantee improvement after each iteration.

\begin{figure}[t]
	\centering
	\subfigure[First Stage]{
		\label{fig:hI_sector}
		\includegraphics[width=0.4\linewidth]{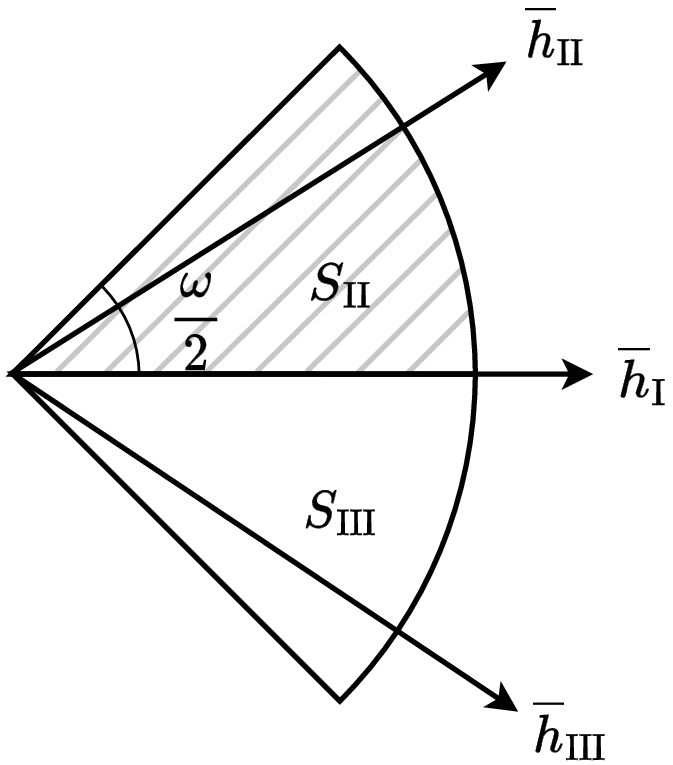}}
	\qquad
	\subfigure[Second Stage]{
		\label{fig:hII_sector}
		\includegraphics[width=0.35\linewidth]{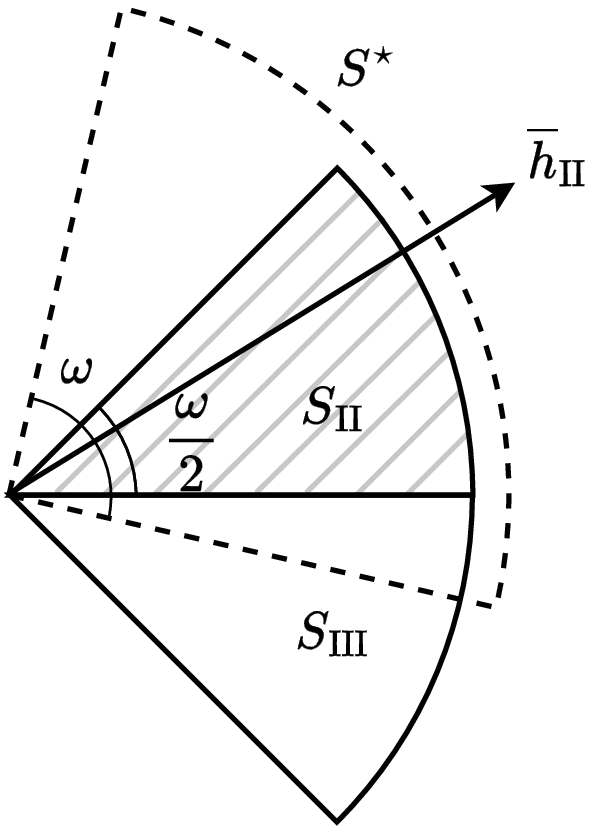}}
	\caption{Visualization of the procedure of Algorithm \ref{alg:GCSM}.}
	\label{fig:sector}
\end{figure}

\begin{algorithm}[t]
\caption{Grouped Conditional Sample Mean (GCSM)}
\label{alg:GCSM}
\begin{algorithmic}[1]
    \State\textbf{input:} $\Phi_K$ and $N$.
    \State Divide the reflected channels into two groups $\mathcal S_{\mathrm{I}}$ and $\mathcal S^c_{\mathrm{I}}$.
    \For{$t=1,2,\ldots,T_1$}
    \State Fix $\mathcal S_{\mathrm{I}}$ and generate each $\theta_{nt}$ with $n\in\mathcal S^c_{\mathrm{I}}$ at random.
    \State Measure received signal power $\left|Y_{t}\right|^2$ under $\boldsymbol{\theta}_t$.
    \EndFor
    \State \textbf{1st round of CSM:} Compute $\widehat{\mathbb{E}}\left[|Y|^2 \mid \theta_n=k \omega\right]$ in \eqref{eq:conditional_sample_mean} and decide $\theta_n$ for each RE in $\mathcal S^c_{\mathrm{I}}$ as in \eqref{eq:theta_selection_CSM}.
    \State Further divide $\mathcal S^c_{\mathrm{I}}$ into $\mathcal S_{\mathrm{II}}$ and $\mathcal S_{\mathrm{III}}$ according to \eqref{find_S}.
    \For{$t=1,2,\ldots,T_2$}
    \State Fix $\mathcal S_{\mathrm{II}}$ and generate each $\theta_{nt}$, $n\in \mathcal S_{\mathrm{I}}\cup \mathcal S_{\mathrm{III}}$ at random.
    \State Measure received signal power $\left|Y_{t}\right|^2$ under $\boldsymbol{\theta}_t$.
    \EndFor
    \State \textbf{2nd round of CSM:} Compute $\widehat{\mathbb{E}}\left[|Y|^2 \mid \theta_n=k \omega\right]$ in \eqref{eq:conditional_sample_mean} and decide $\theta_n$ for each RE in $\mathcal S_{\mathrm{I}}$ and $S_{\mathrm{III}}$ as in (\ref{eq:theta_selection_CSM});
\end{algorithmic}
\end{algorithm} 

Rather interestingly, it turns out that the performance can be guaranteed if the REs are divided into three groups, as explained intuitively in what follows. First, divide the $N$ REs randomly into two groups $\mathcal{S}_{\mathrm{I}}$ and $\mathcal{S}^c_{\mathrm{I}}$. Fixing the phase shifts of the REs in $\mathcal{S}_{\mathrm{I}}$, we optimize $\theta_n$ for each RE $n\in\mathcal{S}^c_{\mathrm{I}}$ by CSM. In other words, the reflected channels associated with $\mathcal{S}_{\mathrm{I}}$ are currently combined with the original direct channel $h_0$ to form a new direct channel denoted by $h_{\mathrm{I}}$; we write $\mathbb E[h_{\mathrm{I}}]$ as $\overline{h}_{\mathrm{I}}$. The resulting $h_{\mathrm{I}}$ would be sufficiently strong, so the CSM algorithm now works for optimizing the phase shifts of the REs in $\mathcal{S}^c_{\mathrm{I}}$. Thus, according to the former discussion in Section \ref{subsec:CSM fading}, $\overline{h}_n$ of each RE in $\mathcal{S}^c_{\mathrm{I}}$ would be rotated by CSM to the closest possible position to $\overline{h}_\mathrm{I}$. In particular, notice that the rotated $\overline{h}_n$, which is $\overline{h}_{n}e^{j\theta_n}$, must lie in two sectors (each of angel $\omega/2$) adjacent to $\overline{h}_{\mathrm{I}}$, as illustrated in Fig.~\ref{fig:hI_sector}. The above steps are referred to as the first stage of our proposed algorithm.

We now enter the second stage of the proposed algorithm. To start with, we further split $\mathcal S_{\mathrm{I}}^c$ into two subgroups. Recall that $\overline{h}_ne^{j\theta_n}$ of each RE $n\in\mathcal S_{\mathrm{I}}^c$ must lie in either the upper sector or the lower sector adjacent to $\overline{h}_{\mathrm{I}}$. As illustrated in Fig.~\ref{fig:hII_sector}, for each RE $n\in\mathcal S_{\mathrm{I}}^c$, we put it in group $\mathcal{S}_{\mathrm{II}}$ if its current $\overline{h}_ne^{j\theta_n}$ lies in the upper sector, and put it in group $\mathcal{S}_{\mathrm{III}}$ if its current $\overline{h}_ne^{j\theta_n}$ lies in the lower sector. Nevertheless, since $\{h_n\}$ are unknown, we do not know the positions of $\{h_ne^{j\theta_n}\}$ either, so how do we determine $\mathcal S_{\mathrm{II}}$ and $\mathcal S_{\mathrm{III}}$ in practice? To find the answer, the key observation is that $\overline{h}_n e^{j\left(\theta_n-\omega\right)}$ is closer to $\overline{h}_{\mathrm{I}}$ than $\overline{h}_n e^{j\left(\theta_n+\omega\right)}$ is if $\overline{h}_n e^{j\theta_n}$ lies in the upper sector, so $\widehat{\mathbb{E}}\left[|Y|^2 \mid \theta_n=\varphi-\omega\right]>\widehat{\mathbb{E}}\left[|Y|^2 \mid \theta_n=\varphi+\omega\right]$. Likewise, we would have $\widehat{\mathbb{E}}\left[|Y|^2 \mid \theta_n=\varphi-\omega\right]<\widehat{\mathbb{E}}\left[|Y|^2 \mid \theta_n=\varphi+\omega\right]$ if $\overline{h}_n e^{j\theta_n}$ lies in the lower sector. Hence, we can decide the group for each $n\in\mathcal S_{\mathrm{I}}^c$ as
\begin{equation}
    \label{find_S}
    \widehat{\mathbb{E}}[|Y|^2|\theta_n=\varphi+\omega] \underset{\mathcal S_{\mathrm{III}}}{\overset{\mathcal S_{\mathrm{II}}}{\lessgtr}} \widehat{\mathbb{E}}[|Y|^2|\theta_n=\varphi-\omega].
\end{equation}
After deciding the groups $\mathcal{S}_{\mathrm{II}}$ and $\mathcal{S}_{\mathrm{III}}$, we further denote by $h_{\mathrm{II}}$ the superposition of all those $h_n e^{j\theta_n}$'s associated with $\mathcal{S}_{\mathrm{II}}$, and write $\overline{h}_{\mathrm{II}}=\mathbb E[h_{\mathrm{II}}]$. Next, we fix the phase shifts of those REs in $\mathcal S_{\mathrm{II}}$, and optimize the phase shifts for the rest REs (which are contained in $\mathcal{S}_{\mathrm{I}}\cup\mathcal{S}_{\mathrm{III}}$) by the CSM algorithm. Thus, during the process of CSM this time, we treat the combination of $h_{\mathrm{II}}$ and $h_0$ as the new virtual direct channel. The second stage is then completed. The whole algorithm now ends. We summarize the above steps in Algorithm \ref{alg:GCSM}. The computational complexity of the proposed algorithm is $O(N(T+K))$. Algorithm \ref{alg:GCSM} has provable performance, as stated in the following proposition.

\begin{proposition}
For any fixed $\xi\in(0,1)$, the solution $\bm\theta^{\mathrm{GCSM}}$ by Algorithm \ref{alg:GCSM} with $T_1=\Omega(N^3\log N)$ and $T_2=\Omega(N^3\log N)$ satisfies
\begin{equation}
    \left|\angle 
    \overline{h}_n+\theta_n^{\mathrm{GCSM}}-\angle \overline{h}_{n'}-\theta_{n'}^{\mathrm{GCSM}}\right| \leq \omega
    \label{GCSM:sector_bound}
\end{equation}
for any two $n,n'\in\{1,2,\ldots,N\}$ with a probability of at least $(1-\xi)^2$, i.e., the LoS components of reflected channels are clustered within a sector of angle $\omega$, so Algorithm \ref{alg:GCSM} can guarantee that
\begin{equation}
\label{GCSM:bounds}
(1-\xi)^2\cos^2(\frac{\pi}{K})\cdot f^\star\le\mathbb{E}\left[\bigg|h_0+\sum_{n=1}^{N}h_ne^{j\theta_n^{\mathrm{GCSM}}}\bigg|^2\right]\le f^\star,
\end{equation}
where $f^\star$ represents the global optimum of problem \eqref{opt problem}.
\end{proposition}
\begin{IEEEproof}
We start by reviewing the procedure of Algorithm \ref{alg:GCSM} briefly. The algorithm comprises two stages. At the first stage, we fix $\theta_n$ for those $h_n\in\mathcal{S}_{\mathrm{I}}$, and optimize $\theta_n$ for those $h_n\in\mathcal{S}_{\mathrm{I}}^c$ by the CSM algorithm. Note that $T_1=\Omega(N^3\log N)$ suggests $T_1=\Omega(|\mathcal{S}_{\mathrm{I}}^c|^3\log |\mathcal{S}_{\mathrm{I}}^c|)$,so the CSM algorithm would rotate $\overline{h}_n$ of each $h_n$ in $ \mathcal{S}_{\mathrm{I}}^c$ to the closest possible position to $\overline{h}_{\mathrm{I}}$ with a probability of at least $1-\xi$ according to Proposition \ref{prop:equivalent_fading}; Fig.~\ref{fig:hI_sector} illustrates the result of the first stage. Subsequently, at the second stage, we fix $\theta_n$ for those $h_n\in\mathcal{S}_{\mathrm{II}}$, and optimize $\theta_n$ for those $h_n\in\mathcal{S}_{\mathrm{I}}\cup\mathcal{S}_{\mathrm{III}}$ by the CSM algorithm. Similarly, with $T_2=\Omega(N^3\log N)$ suggesting $T_2=\Omega(|\mathcal{S}_{\mathrm{I}}\cup\mathcal{S}_{\mathrm{III}}|^3\log |\mathcal{S}_{\mathrm{I}}\cup\mathcal{S}_{\mathrm{III}}|)$, Proposition \ref{prop:equivalent_fading} guarantees that the CSM algorithm would rotate $\overline{h}_n$ of each $h_n\in\mathcal{S}_{\mathrm{I}}\cup\mathcal{S}_{\mathrm{III}}$ to the closest possible position to $\overline{h}_{\mathrm{II}}$ with a probability of at least $1-\xi$, as illustrated in  Fig.~\ref{fig:hII_sector}. Combining the results of two stages, we can guarantee that all the LoS components $\overline{h}_n$ of the reflected channels are rotated to within a section of angle $\omega$---which is denoted by the dashed sector in Fig.~\ref{fig:hII_sector}, with a probability of at least $(1-\xi)^2$. We then establish \eqref{GCSM:sector_bound}. Equipped with \eqref{GCSM:sector_bound}, we can readily obtain \eqref{GCSM:bounds} by following the proof of Proposition \ref{prop:CSM fading}.
\end{IEEEproof}

% \begin{remark}
% \textcolor{blue}{
%     For the binary beamforming case, i.e., when each $\theta_n\in \{0, \pi\}$, the lower bound equals zero in the worst-case scenario, which can be understood through Example \ref{example:CPP_K2}. But we emphasize that Example \ref{example:CPP_K2} is a crafted case. In most random realizations of the $K=2$ case, the proposed algorithm GCSM still performs quite well, as shown in Section \ref{sec:Numerical}.}
% \end{remark}

\begin{remark}[Why not divide RE into more groups?]
    First of all, we would like to clarify that the purpose of dividing the REs into groups is to address the NLoS issue, but it sacrifices the RE coordination. In other words, if the LoS channel is already sufficiently strong, then it is better to put all the REs in the same group. This can be seen from the optimization viewpoint: when maximizing $f(\theta_1,\ldots,\theta_N)$, the optimal method is to optimize all the variables $(\theta_1,\ldots,\theta_N)$ simultaneously, rather than individually, otherwise it is very likely to get stuck at a premature local optimum. Since using more groups is not a good thing for the RE coordination, why don't we just use two groups? The problem with using two groups is that the resulting GCSM algorithm cannot guarantee convergence anymore as shown later in Fig.~\ref{fig:alternative_CSM} in Section \ref{sec:Numerical}. In a nutshell, we suggest dividing the REs into three groups because this is the smallest number of groups that ensures the convergence of GCSM. We use simulations to verify the above argument in Section \ref{sec:Numerical}.
\end{remark}

\begin{figure*}[t]
    \centering    \includegraphics[width=10cm]{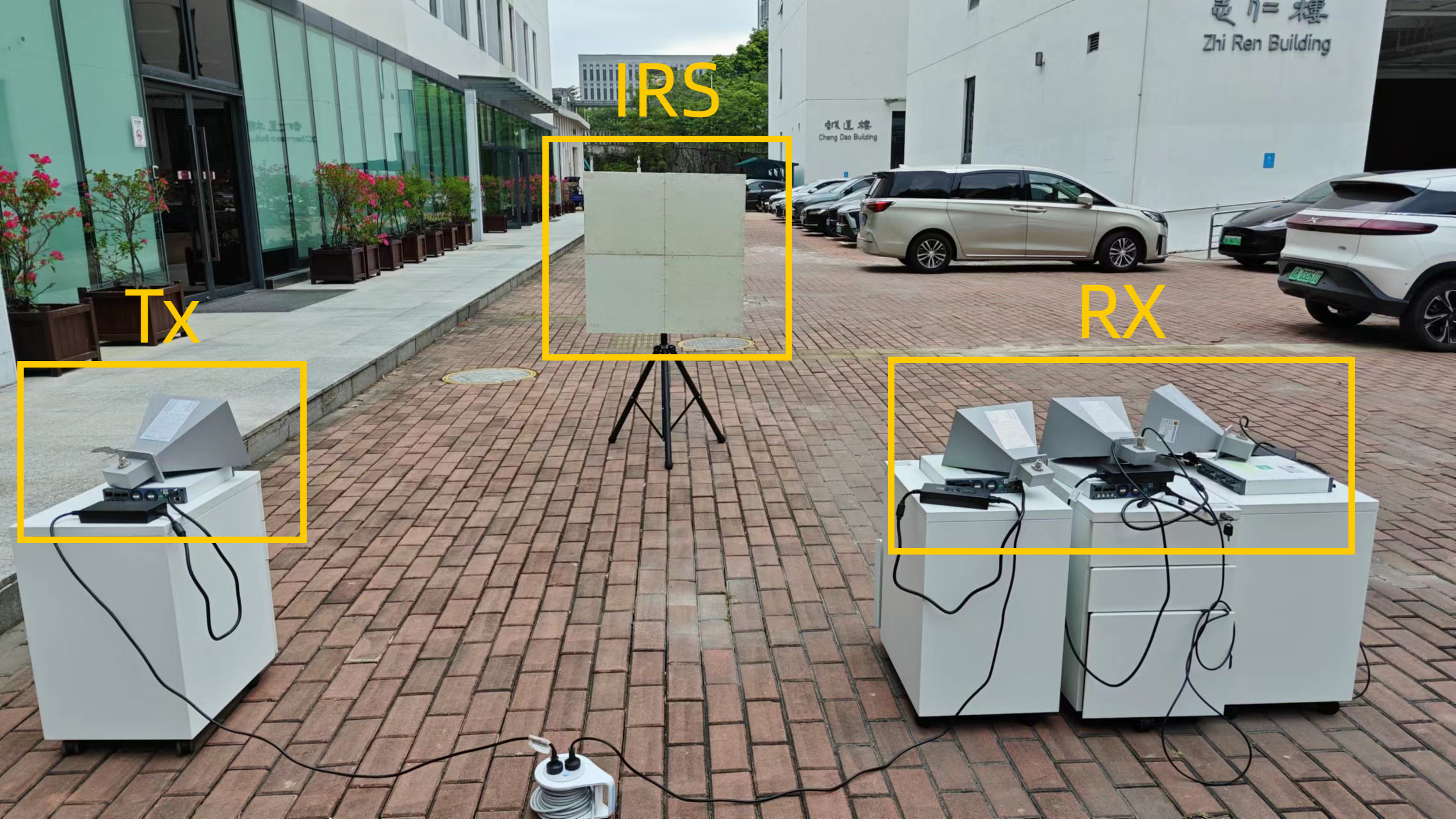}
    \caption{Our field test uses an IS that comprises 400 REs and provides 4 phase shift options $\{0,\pi/2,\pi,3\pi/2\}$ for each RE. For the LoS case, omnidirectional antennas are deployed at all devices; for the NLoS case, directional antennas are used to prevent direct signal propagation from the transmitter to each receiver.}
    \label{fig:outdoor_test}
\end{figure*}

\begin{figure*}[t]
	\centering
	\includegraphics[width=11cm]{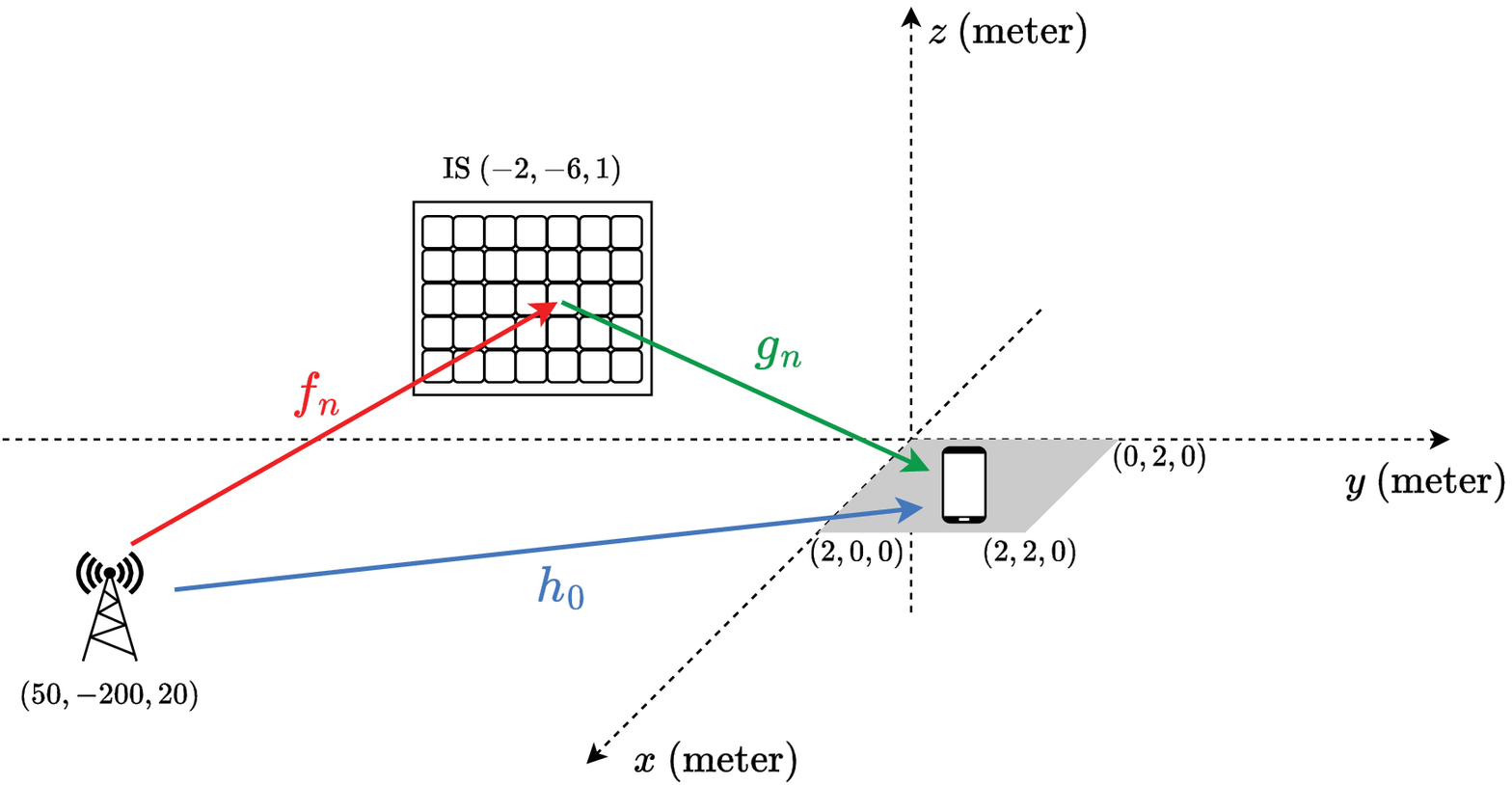}
	\caption{The IS-assisted downlink network considered in our simulations. One or more receivers are randomly located within the shaded area.}
	\label{fig:simulation_loc}
\end{figure*}

\subsection{Extension to Multi-User Case}
\label{sec:multi_user}
We further propose an extension of blind beamforming for multiple users. Consider an IS-assisted broadcast network \cite{yan2023passive} where a single-antenna transmitter sends a common message to $M\geq 2$ single-antenna receivers. We denote by $h_0^m$ the direct channel from the transmitter to the $m$th user, and denote by $h_n^m$ the reflected channel associated with the $m$th user and the $n$th RE. Rician fading is assumed as in \eqref{Rician channel model}. Let $\sigma^2_m$ be the background noise power level at the $m$th user. We seek the optimal passive beamformer $\bm \theta=(\theta_1, \ldots, \theta_N)$ that maximizes the worst SNR across the users, i.e.,
\begin{subequations}
\label{multi opt problem}
\begin{align}
    &\underset{\bm \theta}{\text{maximize}}\quad \underset{m}{\text{min}}\left\{\mathbb{E}\left[\frac{P}{\sigma_m^2}\left|h_0^m+\sum_{n=1}^{N}h_n^m e^{j\theta_n}\right|^2\right] \right\}
    \label{multi opt problem:obj}\\
    &\text {subject to} \quad \theta_n \in \Phi_K,\quad n=1, \ldots, N,
    \label{multi opt problem:contraint}
\end{align}
\end{subequations}
where the expectation is taken over the fading channels.

To account for multiple users, we use the following utility in place of the received signal power for blind beamforming:
\begin{equation}
	\label{eq:max-min utility}
	U_t=\min_m\left\{\left|Y_t^m\right|^2/\sigma_m^2\right\}.
\end{equation}
Accordingly, the empirical conditional average is now computed as
\begin{equation}
	\widehat{\mathbb{E}}\left[U \mid \theta_n=k \omega\right]=\frac{1}{\left|\mathcal{Q}_{n k}\right|} \sum_{t \in \mathcal{Q}_{n k}}U_t.
\end{equation}
For the utility-based CSM, each $\theta_n$ is chosen to maximize the empirical conditional average of the utility:
\begin{equation}
    \theta_n^{\mathrm{CSM}}=\arg \max _{\varphi \in \Phi_K} \widehat{\mathbb{E}}\left[U \mid \theta_n=\varphi\right].
\end{equation}
The utility-based GCSM for multiple users can be obtained similarly.

\section{Numerical results}
\label{sec:Numerical}

\subsection{Field Tests}
We carry out the field test in an outdoor environment as shown in Fig.~\ref{fig:outdoor_test}. Throughout the field test, the transmit power is fixed to be $-10$ dBm and the transmission takes place at $3.5$ GHz. The spectrum bandwidth equals 125 KHz, and the modulation scheme is quadrature amplitude modulation (QAM). The SNR is measured 100 times and then averaged out. The IS prototype machine consists of $400$ REs (i.e., $N=400$) and provides 4 possible phase shifts $\{0,\pi/2,\pi,3\pi/2\}$ for each RE. For LoS direct channel case, omnidirectional antennas are deployed at both transmitter and receiver; for NLoS direct channel case, directional antennas are deployed to prevent direct signal propagation from transmitter to receiver. Aside from the CSM algorithm in \cite{ren2022configuring} and the proposed GCSM algorithm in Algorithm \ref{alg:GCSM}, the following methods are included in field tests as benchmarks:
\begin{itemize}
    \item \emph{Without IS:} IS is removed from the network.
    \item \emph{Zero Phase Shifts (ZPS):} Fix all phase shifts to be zero.
    \item \emph{Beam Training \cite{ren2022configuring}:} Try out $T$ random samples of the phase shift array $\bm \theta=(\theta_1,\ldots, \theta_N)$ and choose the best.
\end{itemize}
Notice that beam training, CSM, and GCSM all require random samples of $\boldsymbol{\theta}$. We let $T=1000$ and $T_1=T_2=500$. 

Table~\ref{tab:SNR_boost} summarizes the performance of the different IS beamforming algorithms. Observe that the ZPS method increases SNR significantly by 23.8 dB for the NLoS direct channel case even with the phase shifts all fixed at zero, whereas its gain is marginal (only 1.8 dB) for the LoS case. Observe also that beam training increases SNR further by about 8 dB as compared to ZPS, at the cost of 1000 random samples. Thus, the deployment of IS can already bring considerable gain even without any phase shift optimization if the original channel is too bad.

But the much higher gain can be reaped by using more sophisticated algorithms like CSM and GCSM. The table shows that CSM can further improve upon beam training by around 7 dB in the LoS case; we remark that CSM uses the same number of random samples as beam training does, and that its computational complexity is not higher. Notice that the gap between CSM and GCSM is slim in the LoS case as expected. But when it comes to the NLoS case, GCSM starts to outperform: SNR of GCSM is 2.4 dB higher than that of CSM. We also notice that the advantage of CSM over beam training shrinks in the NLoS case. But since the two algorithms require similar sampling and computation costs, the former is still preferable in practice.

Table \ref{tab:multi_SNR_boost} further compares the performance of the different IS beamforming algorithms in a broadcast network with $M=3$ users as shown in Fig.~\ref{fig:outdoor_test}. The deployment of the IS already increases the worst SNR among three users by 1.3 dB for the LoS direct channel case and by 2.5 dB for the NLoS case. The best performance in the LoS case is achieved by CSM, but it is only slightly better than GCSM. When it comes to the NLoS case, GCSM becomes the best method.

\begin{table}[t]
\small
    \renewcommand{\arraystretch}{1.3}
\centering
\caption{\small SNR Performance (Measured in dB) in the Single-User Case}
\begin{tabular}{lrrrrr}
\firsthline
& \multicolumn{2}{c}{LoS} & \multicolumn{2}{c}{NLoS}\\
\cline{2-3}\cline{4-5}
Method       & SNR & Boost & SNR & Boost  \\
\hline
Without IS & 8.5 & 0.0 & 3.0 & 0.0 \\
ZPS    & 10.3  & 1.8 & 26.8 & 23.8 \\
Beam Training   & 17.7  & 9.2 & 35.7 & 32.7 \\
CSM  & 24.4 & 15.9 & 29.2 & 36.7  \\
GCSM   & 24.8  & 16.3 & 42.1 & 39.1 \\
\lasthline
\end{tabular}
\label{tab:SNR_boost}
\end{table}

\begin{table}[t]
\small
    \renewcommand{\arraystretch}{1.3}
\centering
\caption{\small Worst-SNR Performance (measured in dB) in the Multi-User Case}
\begin{tabular}{lrrrrr}
\firsthline
& \multicolumn{2}{c}{LoS} & \multicolumn{2}{c}{NLoS}\\
\cline{2-3}\cline{4-5}
Method       & SNR & Boost & SNR & Boost  \\
\hline
Without IS  & 4.8 & 0.0 & 3.4 & 0.0 \\
ZPS    & 6.1  & 1.3 & 5.9  & 2.5 \\
Beam Training   & 9.4  & 4.4 & 8.1  & 4.7 \\
CSM  & 11.0 & 6.2 & 8.4  & 5.0 \\
GCSM  & 10.8 & 6.0 & 9.3 & 5.9 \\
\lasthline
\end{tabular}
\label{tab:multi_SNR_boost}
\end{table}

\begin{figure}[t]
    \centering
    \includegraphics[width=9cm]{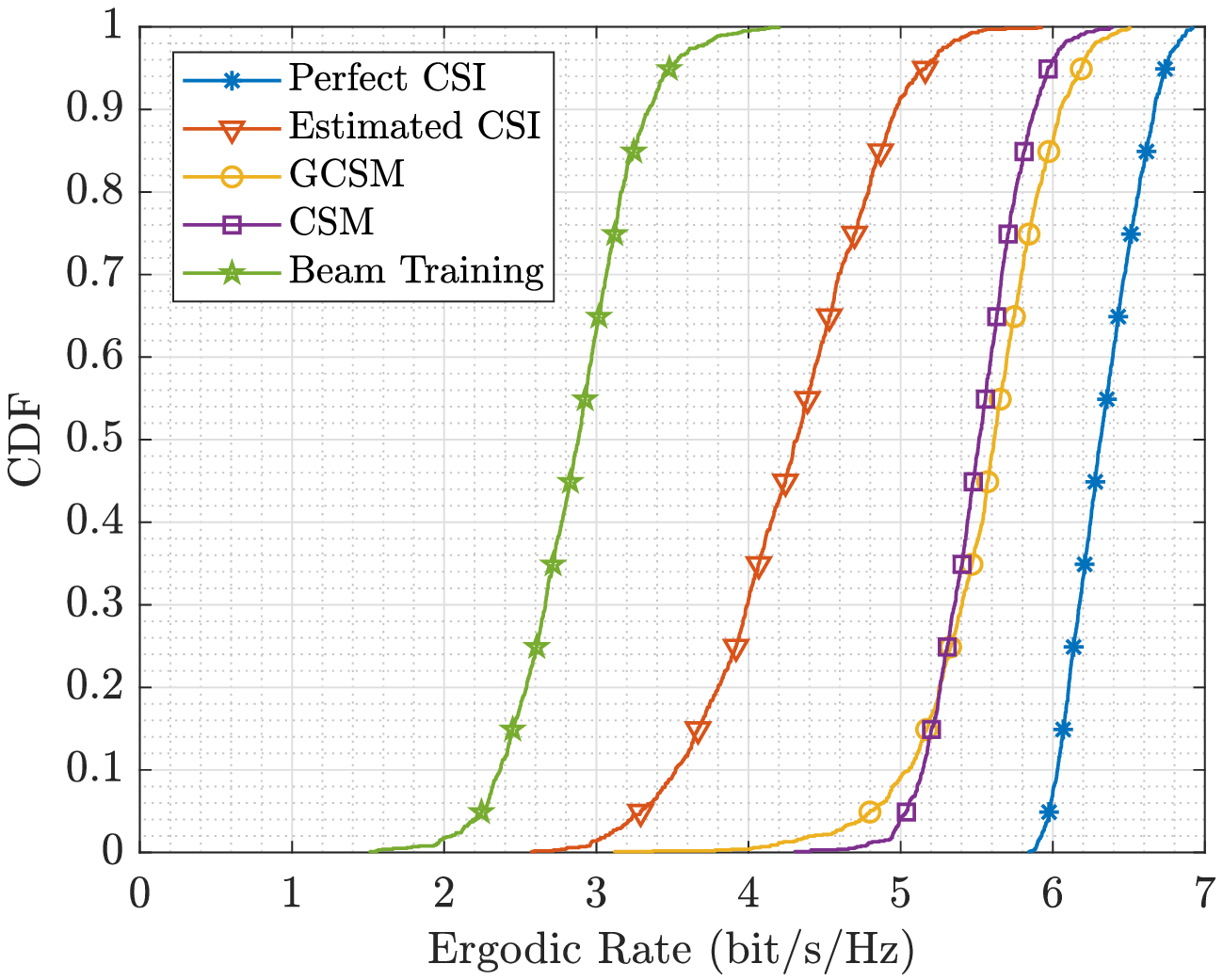}
    \caption{CDF of ergodic rates in the LoS case.}
    \label{fig:SISO_Rician}
\end{figure}
\begin{figure}[t]
    \centering
    \includegraphics[width=9cm]{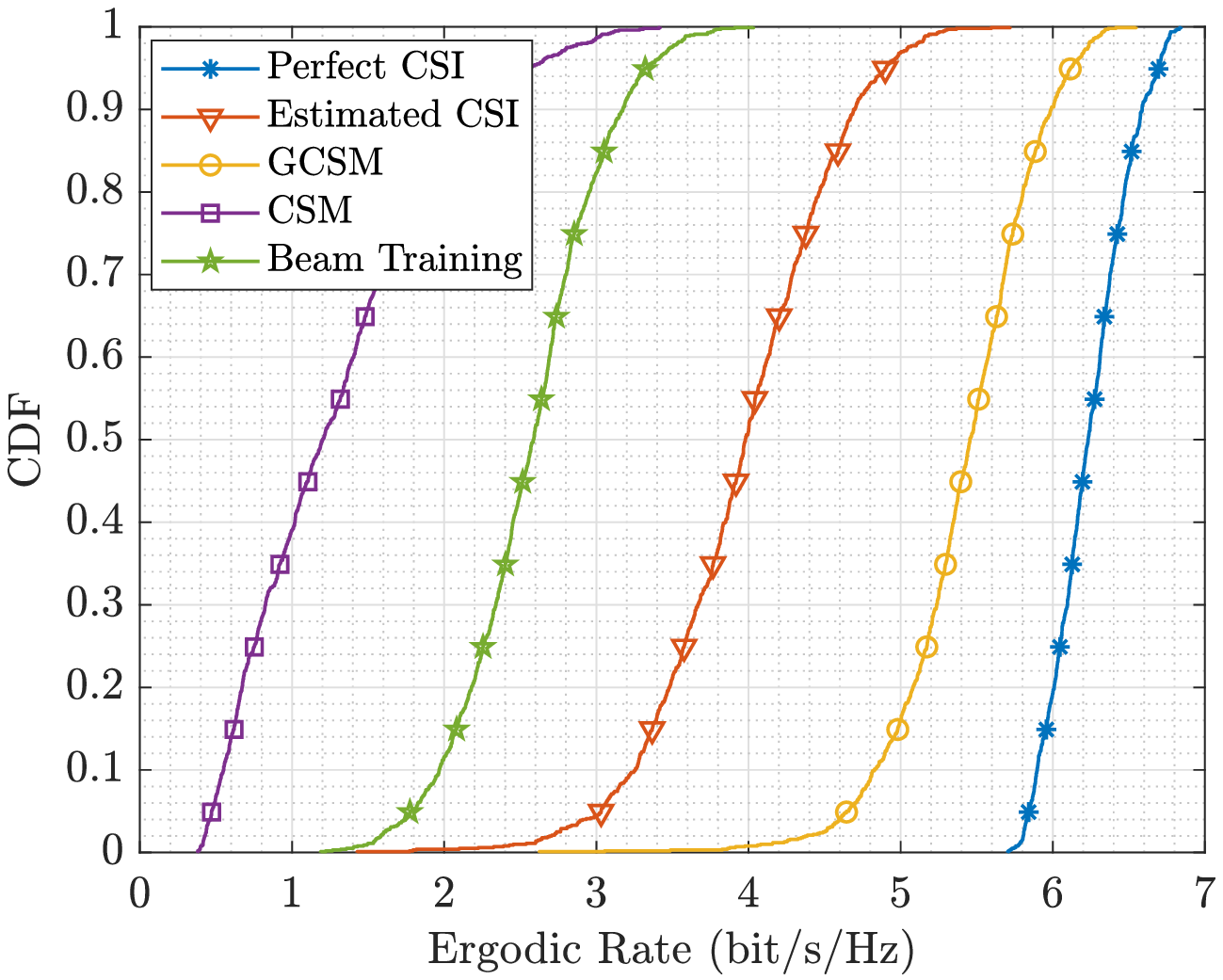}
    \caption{CDF of ergodic rates in the NLoS case.}
    \label{fig:SISO_Rayleigh}
\end{figure}

\begin{figure}[t]
    \centering
    \includegraphics[width=9cm]{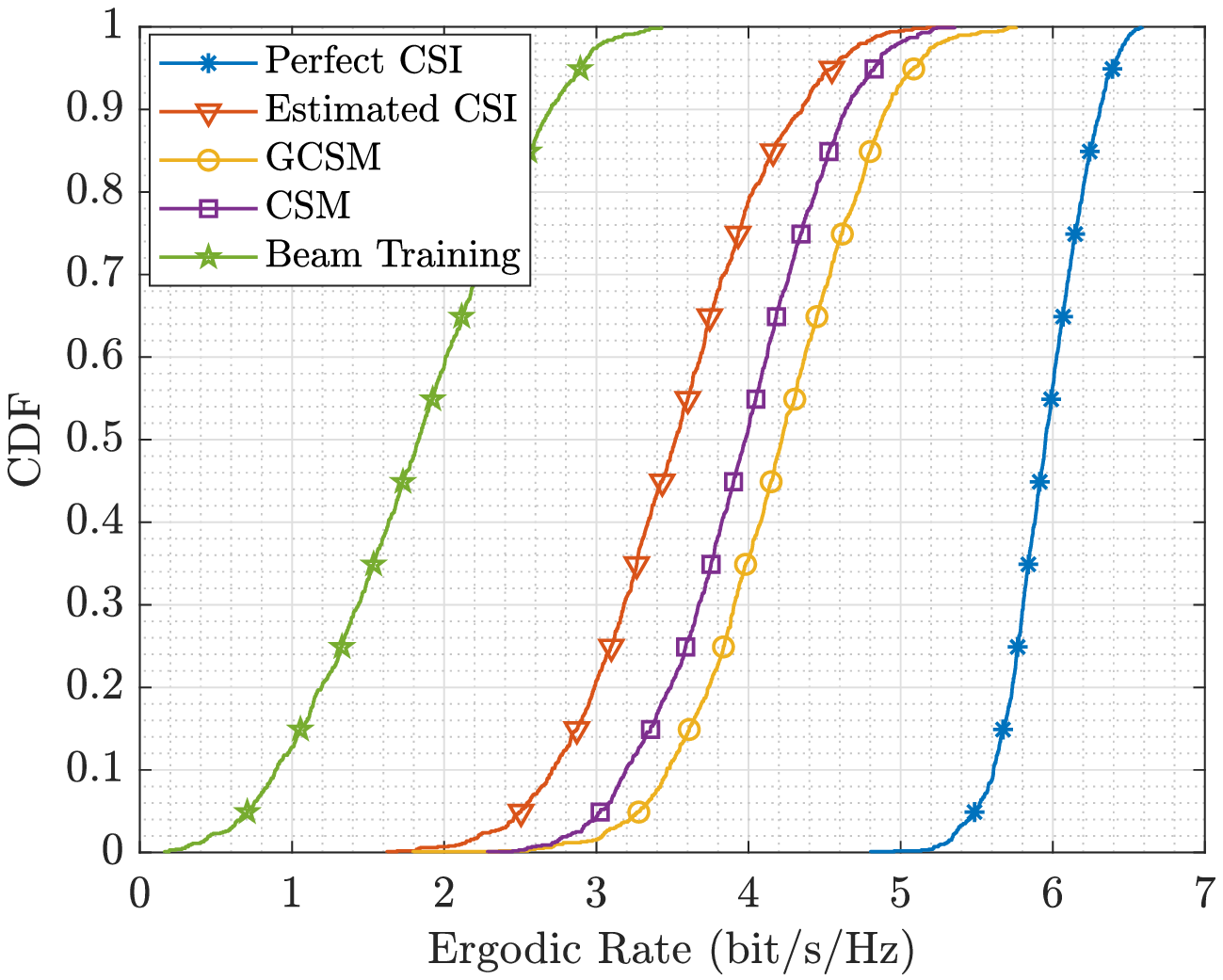}
    \caption{CDF of ergodic rates in the LoS case with interference.}
    \label{fig:Rician_interference}
\end{figure}
\begin{figure}[t]
    \centering
    \includegraphics[width=9cm]{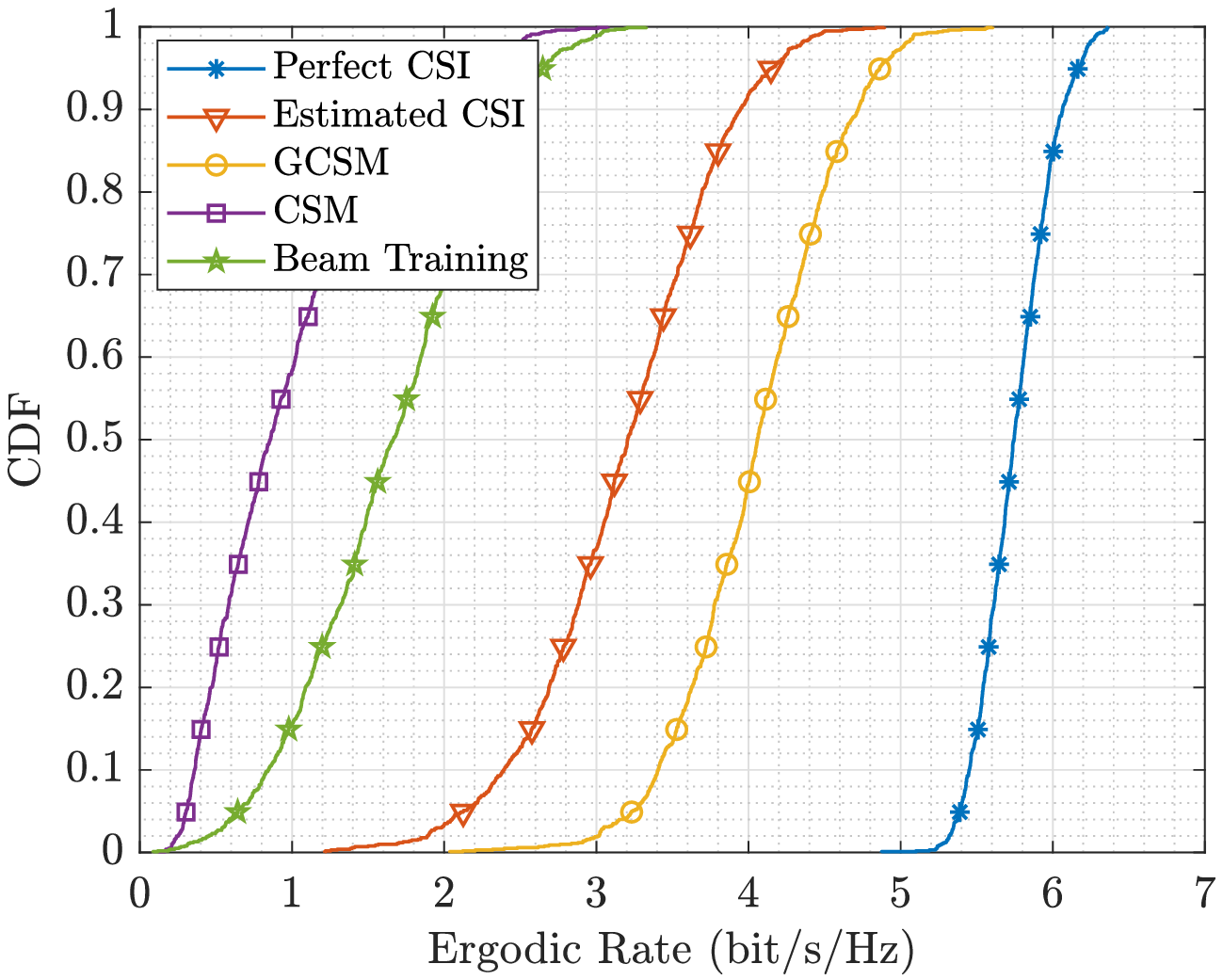}
    \caption{CDF of ergodic rates in the NLoS case with interference.}
    \label{fig:Rayleigh_interference}
\end{figure}

\begin{figure}[t]
    \centering
    \includegraphics[width=9cm]{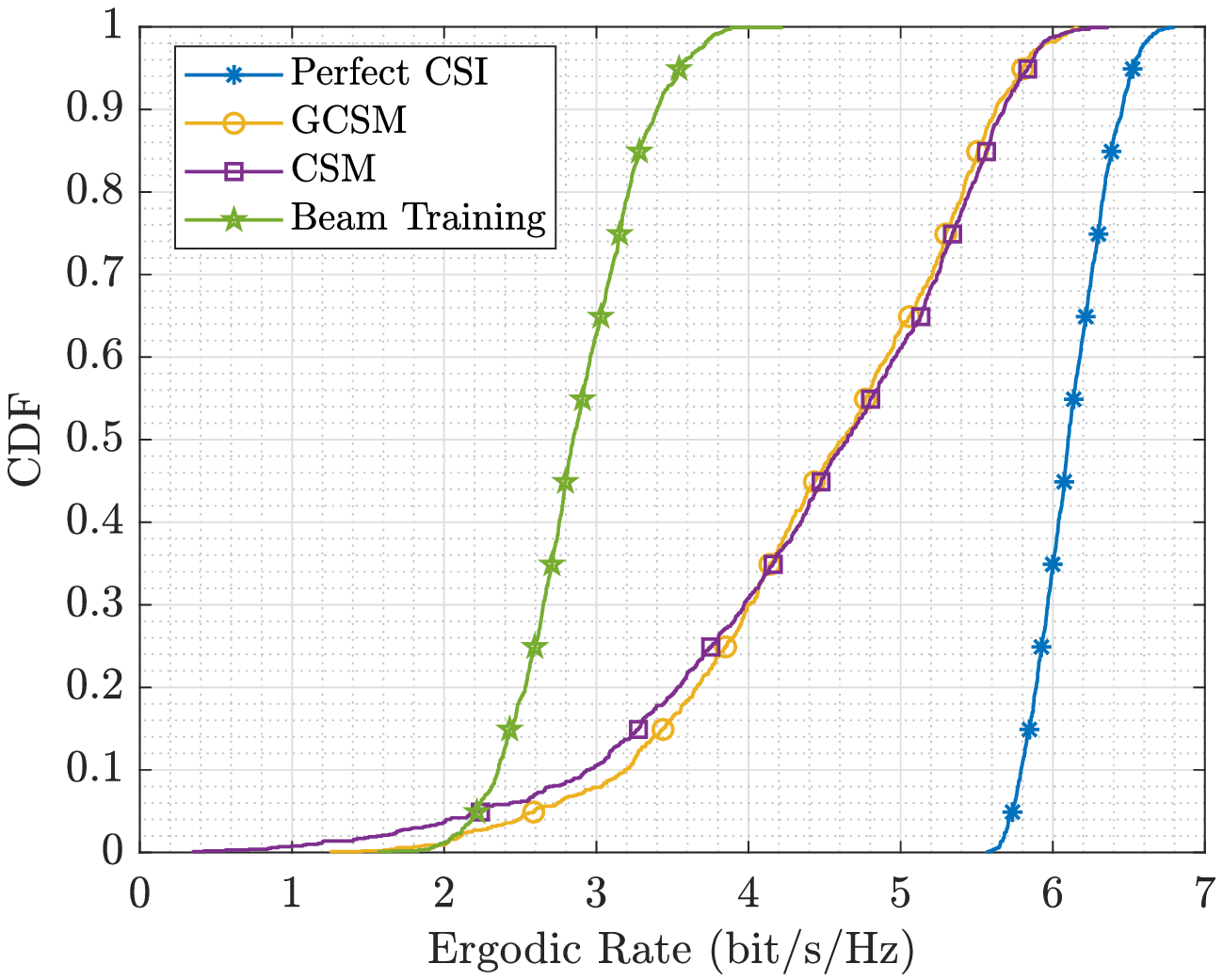}
    \caption{CDF of ergodic rates in the LoS case when $K=2$.}
    \label{fig:Rician_K2}
\end{figure}

\begin{figure}[t]
    \centering
    \includegraphics[width=9cm]{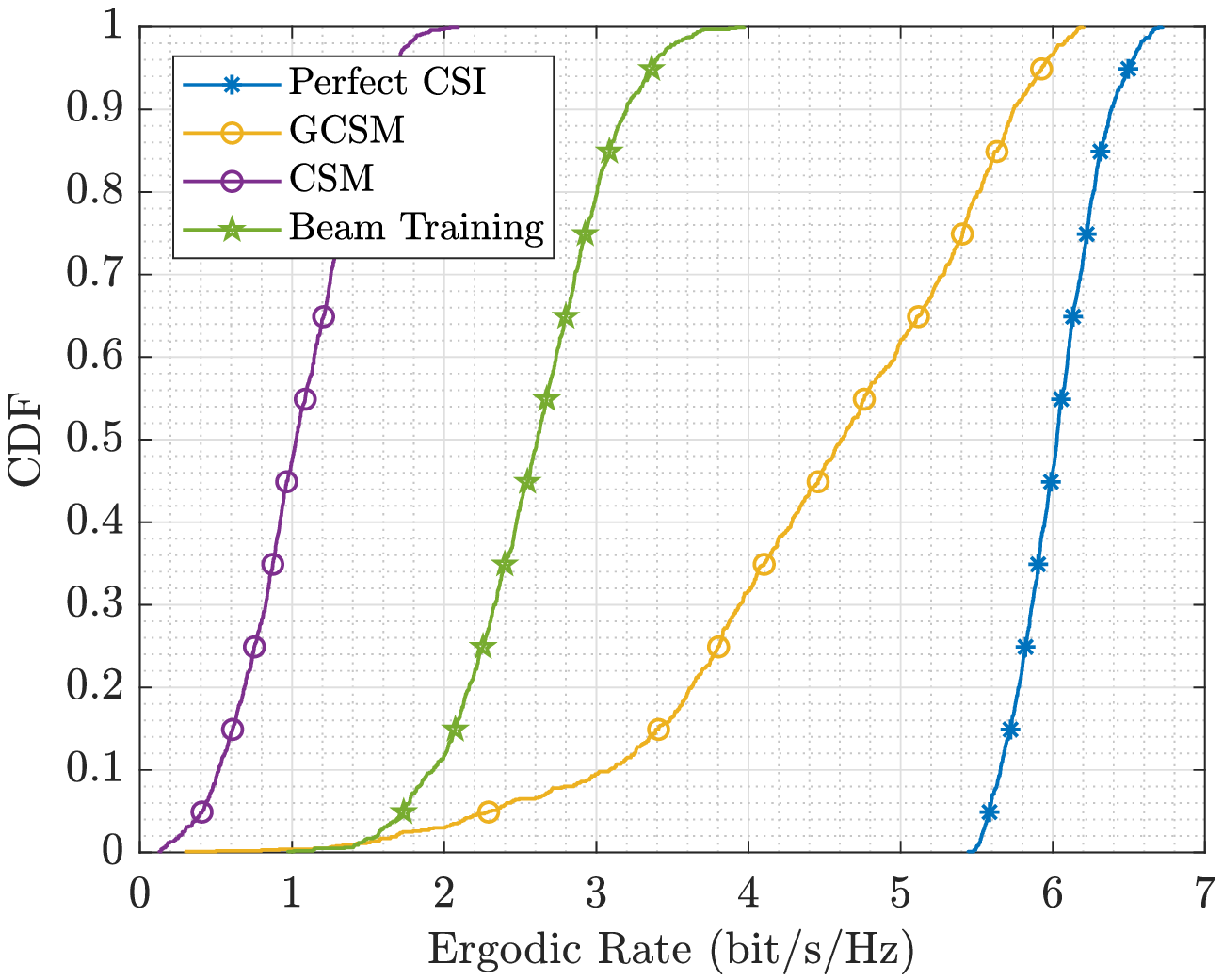}
    \caption{CDF of ergodic rates in the NLoS case when $K=2$.}
    \label{fig:Rayleigh_K2}
\end{figure}

\begin{figure*}[t]
\centering
\subfigure[Perfect CSI]{
\includegraphics[width=0.18\linewidth]{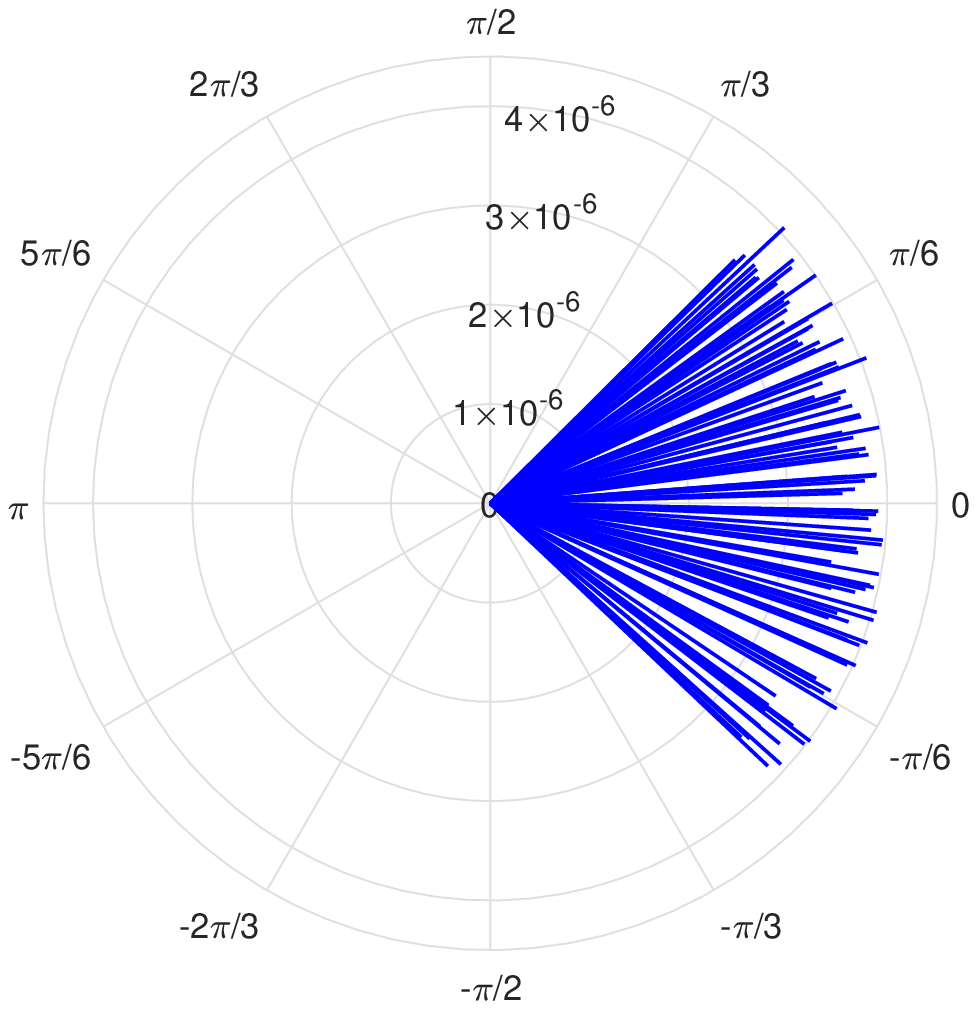}}
\qquad
\subfigure[Estimated CSI]{
\includegraphics[width=0.18\linewidth]{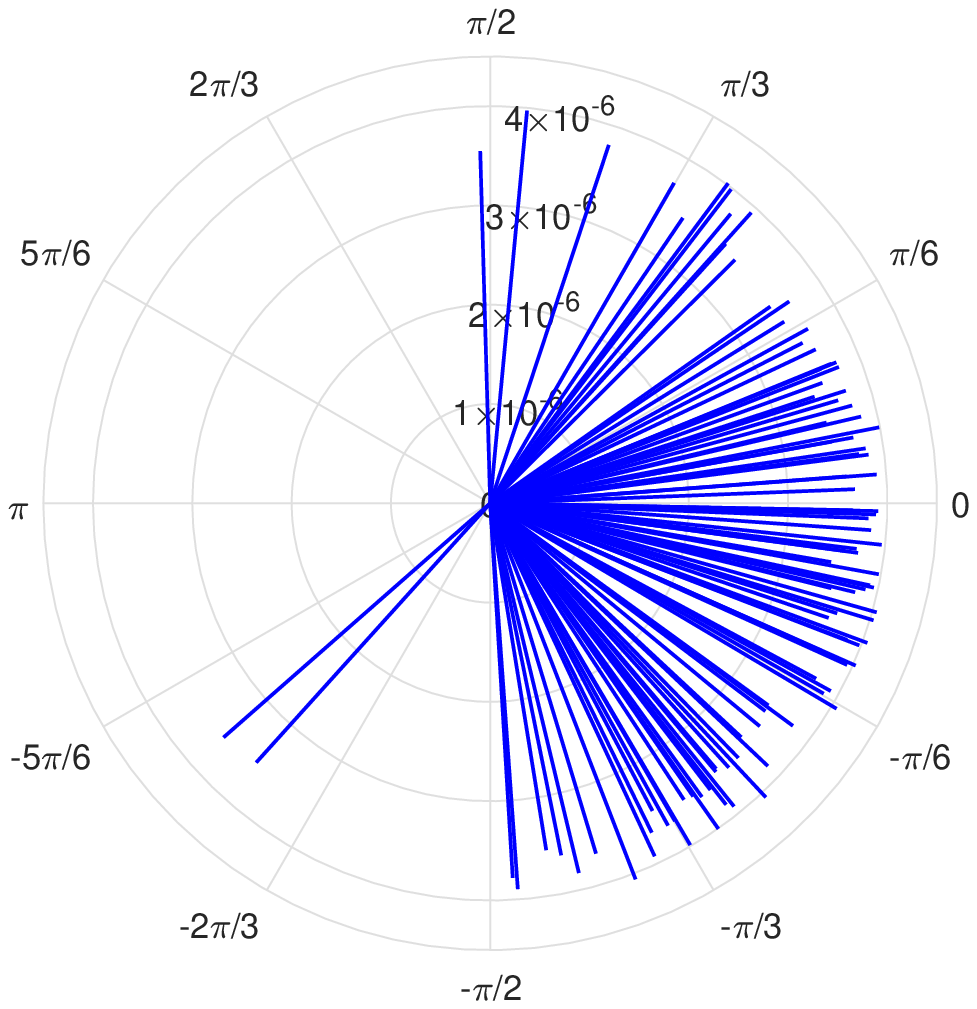}}
\qquad
\subfigure[GCSM]{
\includegraphics[width=0.18\linewidth]{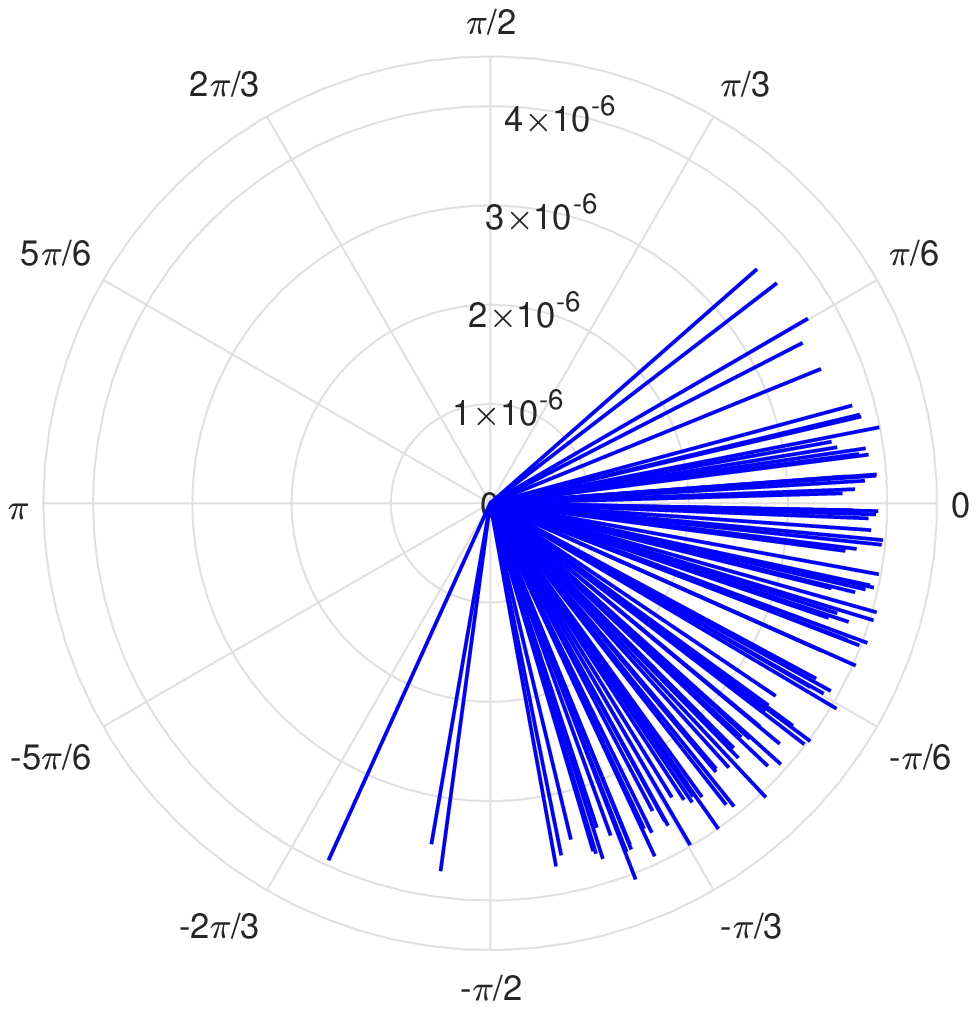}}
\qquad
\subfigure[CSM]{
\includegraphics[width=0.18\linewidth]{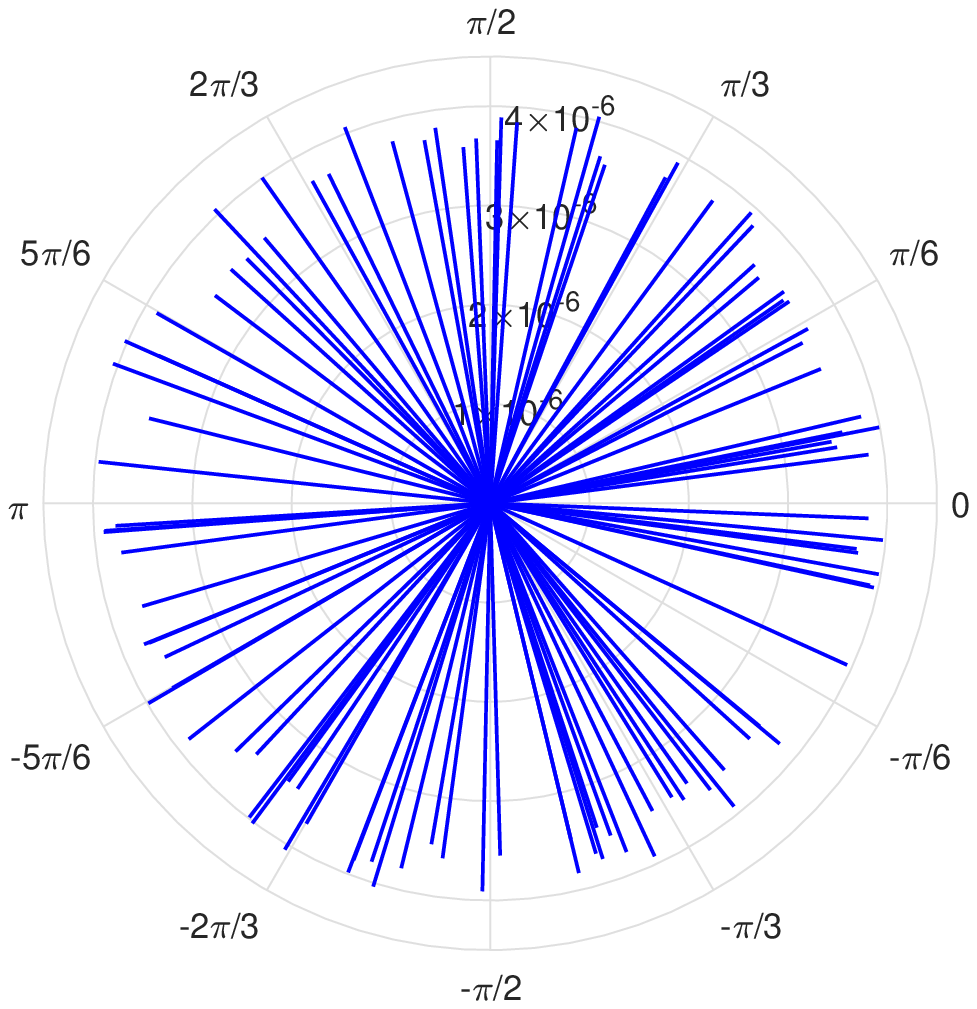}}
\caption{The phase shifted channels $\overline{h}_ne^{j\theta_n}$ in the complex plane by the different IS beamforming methods.}
\label{fig:polar}
\end{figure*}

\subsection{Simulation Tests}
We now consider simulations to validate the performance of the proposed blind beamforming method under more complex network settings, i.e., with more than one user and with many more REs on each IS. Our simulations are conducted on a computer with a 4.6 GHz i5-11500 CPU and 16 GB RAM.

As shown in Fig.~\ref{fig:simulation_loc}, we consider an IS-assisted downlink transmission system. The REs are arranged as a half-wavelength
spaced uniform linear array (ULA); the carrier frequency equals 2.6 GHz, so the wavelength $\lambda\approx10$ cm and thus the
RE spacing equals 5 cm.
We then specify the model parameters in Section \ref{sec:sys}. The pathloss factors follow \cite{jiang2021learning}, which are generated as
\begin{subequations}
\begin{align}
-10\times\log_{10}\gamma_{00}&=32.6+36.7\log_{10}(d_{00})\\
-10\times\log_{10}\gamma_{0n}&=30+22\log_{10}(d_{0n})\\
-10\times\log_{10}\gamma_{n0}&=30+22\log_{10}(d_{n0}),
\end{align}
\end{subequations}
where $d_{00}$, $d_{0n}$, and $d_{n0}$ are the corresponding distance in meters. Moreover, the normalized fixed components are generated as
\begin{subequations}
\begin{align}
\overline{h}_{0}&=\exp\left(-j\frac{2\pi d_{00}}{\lambda}\right) \\
\overline{f}_{n}&=\exp\left(-j\frac{2\pi d_{0n}}{\lambda}\right) \\
\overline{g}_{n}&=\exp\left(-j\frac{2\pi d_{n0}}{\lambda}\right),
\end{align}
\end{subequations}
Regarding the Rician factors, we let $\delta_{0n}=\delta_{n0}=10$; the value of $\delta_{00}$ depends on the direct channel status---let $\delta_{00}=10$ for the LoS case and let $\delta_{00}=0$ for the NLoS case.

The rest parameters are set as follows unless otherwise stated. The transmit power level $P=20$ dBm, and the background noise power level $\sigma^2=-90$ dBm. Assume that the IS has $N=100$ REs; but we will test different values of $N$ later on. The number of phase shift choices $K$ is fixed to be 4. The sample size $T=1000$ by default; but we will change $T$ to see how it impacts the performance of blind beamforming. We evaluate the ergodic data rate by averaging out $10000$ realizations of fading channels.

Aside from CSM, GCSM, and beam training (see the previous subsection), the following baseline methods are considered:
\begin{itemize}
    \item \emph{Perfect CSI:} Perform the closest point projection in \eqref{CPP} with perfectly known CSI; we remark that CSM or GCSM converges to this scheme if $T\rightarrow\infty$.
    \item \emph{Estimated CSI:} Follow the above baseline method except that CSI is estimated by the DFT method \cite{zheng2019intelligent}.
\end{itemize}

Let us begin with the single-user case. The resulting cumulative distributions (CDFs) of ergodic data rates achieved by the different algorithms in the LoS direct channel case are displayed in Fig.~\ref{fig:SISO_Rician}. It can be seen that the proposed blind beamforming method CSM outperforms the estimated CSI method and the beam training method significantly. 
For instance, CSM improves upon beam training by about 30\% and upon GCSM by about 90\% at the 50th percentile. Observe that CSM and GCSM yield similar performance for the LoS case in Fig.~\ref{fig:SISO_Rician}. By contrast, as shown in Fig.~\ref{fig:SISO_Rayleigh}, GCSM is far superior to CSM when the direct channel becomes NLoS. Actually, CSM is the worst among all competitor algorithms in the NLoS case. Fig.~\ref{fig:SISO_Rician} and Fig.~\ref{fig:SISO_Rayleigh} also show that GCSM is quite close to perfect CSI (which amounts to GCSM with infinitely many samples); thus, using merely $T=1000$ samples is good enough in this case.

Moreover, Fig.~\ref{fig:Rician_interference} and Fig.~\ref{fig:Rayleigh_interference} show the performance of different algorithms for the single-user case in the presence of co-channel interference. Specifically, an interfering BS has been placed at the position $(150, -200, 20)$. It can be seen that the achieved ergodic rate of all algorithms decreases in both LoS and NLoS cases compared to the system without interference. It can also be seen that the performance gap between the Estimated CSI and the proposed blind beamforming algorithms narrows. Notably, in the LoS scenario, GCSM exhibits better performance than CSM, indicating that GCSM is more robust in the presence of interference. Fig.~\ref{fig:Rician_K2} and Fig.~\ref{fig:Rayleigh_K2} further show the performance of different algorithms in the binary beamforming case. It can be observed that the proposed algorithm still performs quite well.

It is worthwhile to look into the NLoS case more closely by comparing the phase shift decisions of the different methods. Fig.~\ref{fig:polar} shows how the reflected channels are rotated in the complex plane by the phase shifts of the different methods. The results here are consistent with what we have observed in Fig.~\ref{fig:SISO_Rayleigh}. The perfect CSI method renders the reflected channels most clustered and hence yields the highest SNR boost. GCSM with $T=1000$ also leads to most channels being clustered within an angle of $\pi/2$, albeit a few reflected channels deviate from the main beam because of the limited samples. In comparison, the estimated CSI method results in the reflected channels being more dispersed, and accordingly its performance is indeed worse than the previous two methods. Notably, CSM has the reflected channels uniformly distributed, so it ends up with the worst performance.

\begin{figure}[t]
    \centering
    \includegraphics[width=9cm]{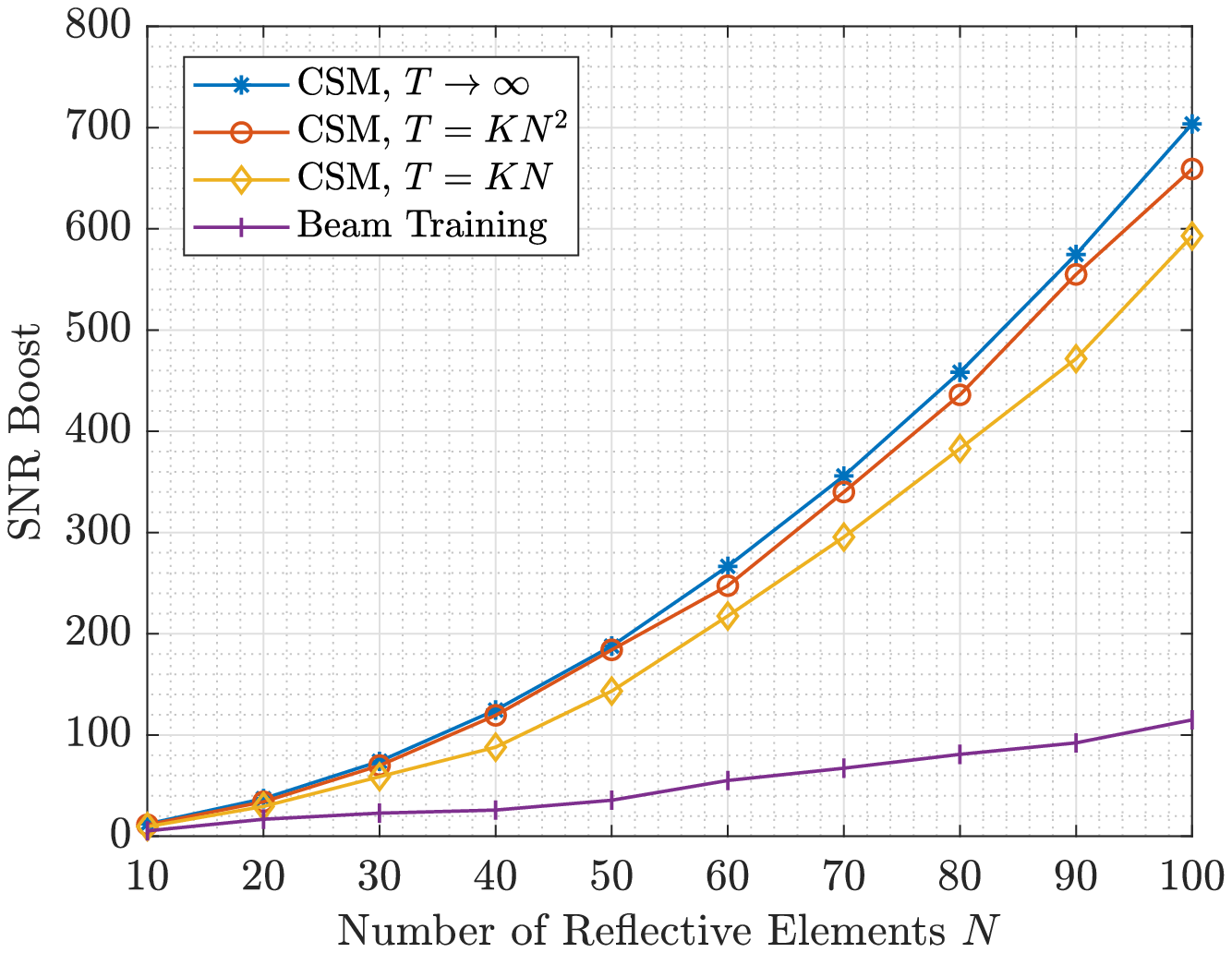}
    \caption{SNR boost vs. $N$ in the LoS case.}
    \label{fig:N2_Rician}
\end{figure}

\begin{figure}[t]
    \centering
    \includegraphics[width=9cm]{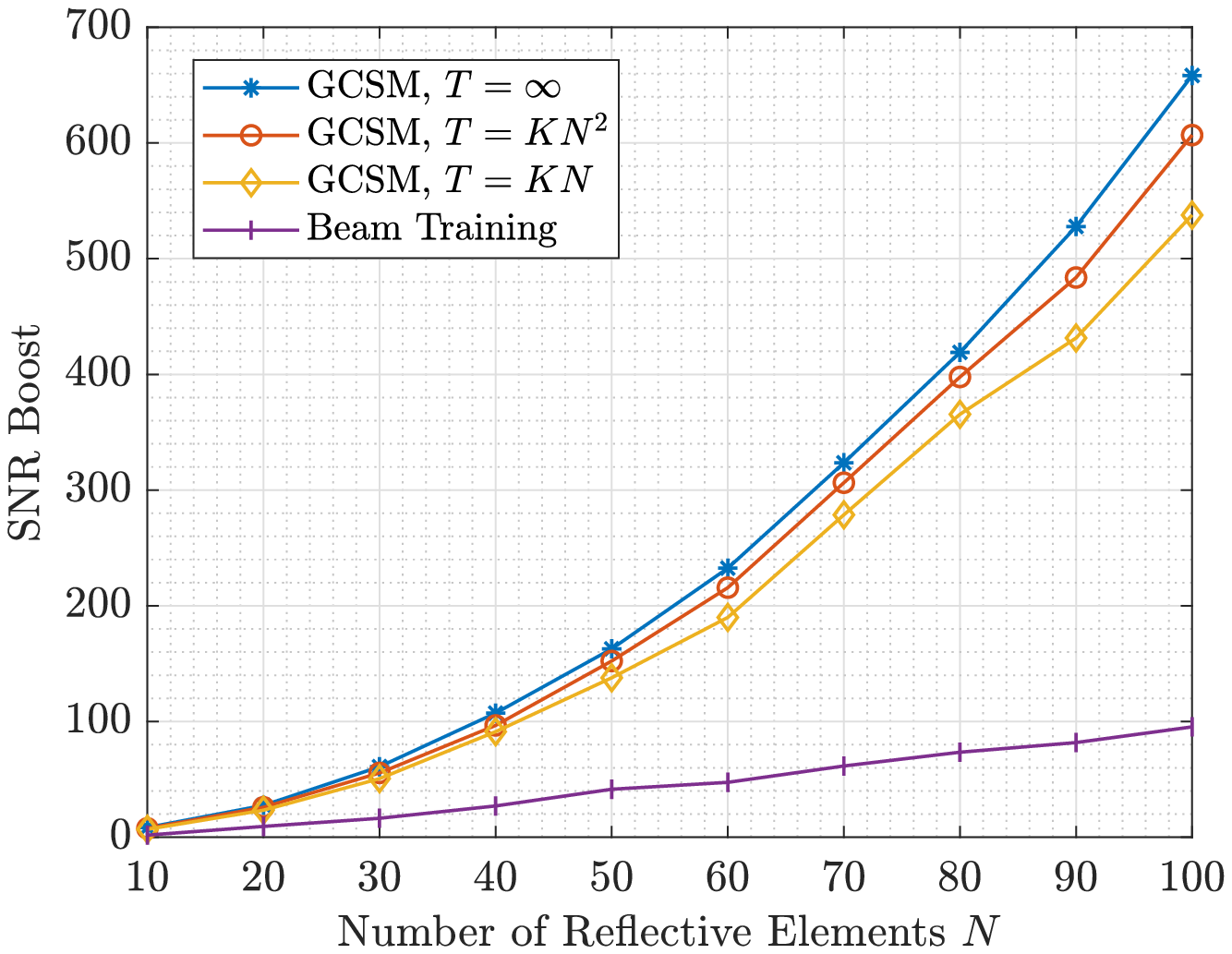}
    \caption{SNR boost vs. $N$ in the NLoS case.}
    \label{fig:N2_Rayleigh}
\end{figure}

\begin{figure}[t]
    \centering
    \includegraphics[width=9cm]{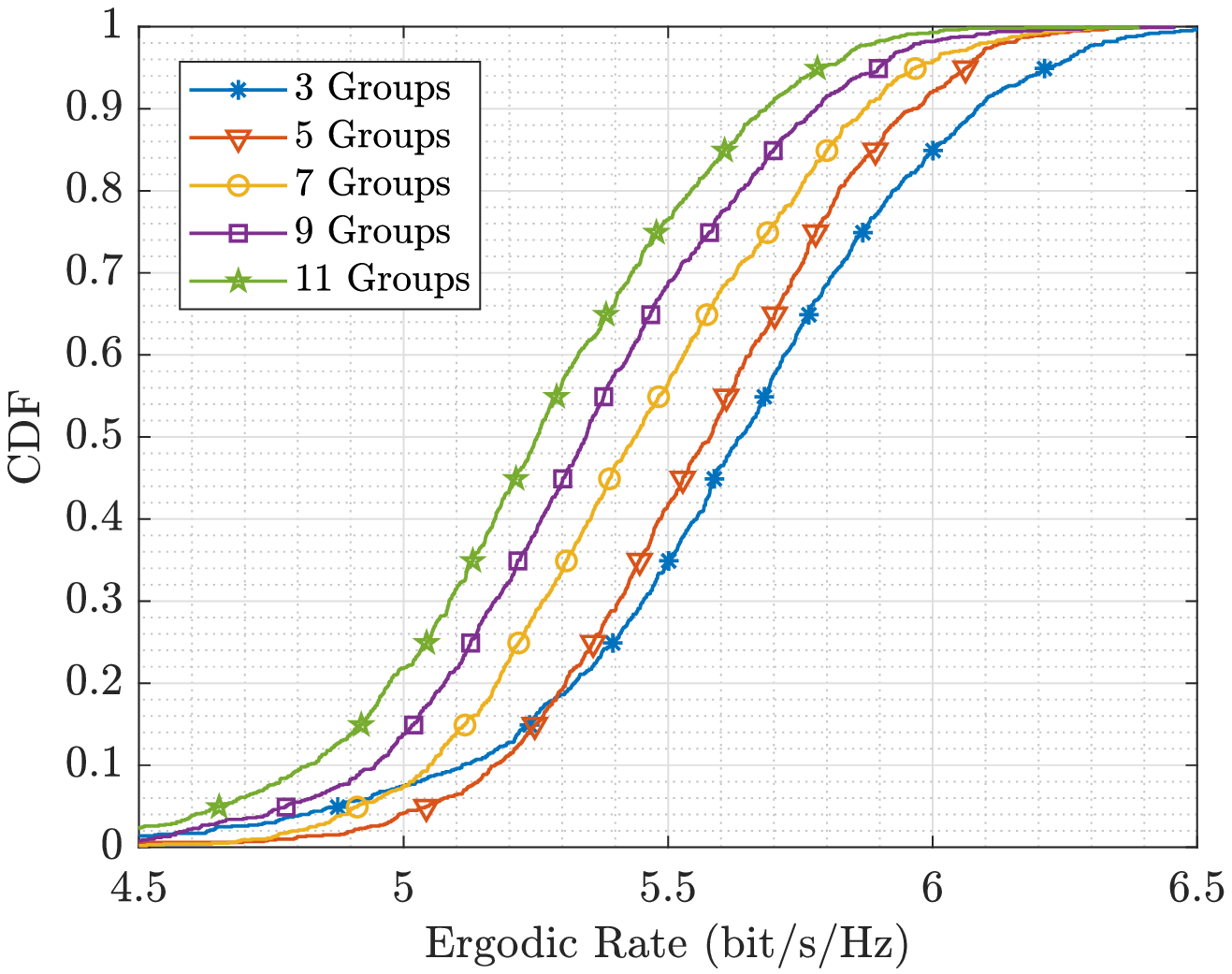}
    \caption{CDF of ergodic rates vs. number of groups in the LoS case.}
    \label{fig:group_LoS}
\end{figure}

\begin{figure}[t]
    \centering
    \includegraphics[width=9cm]{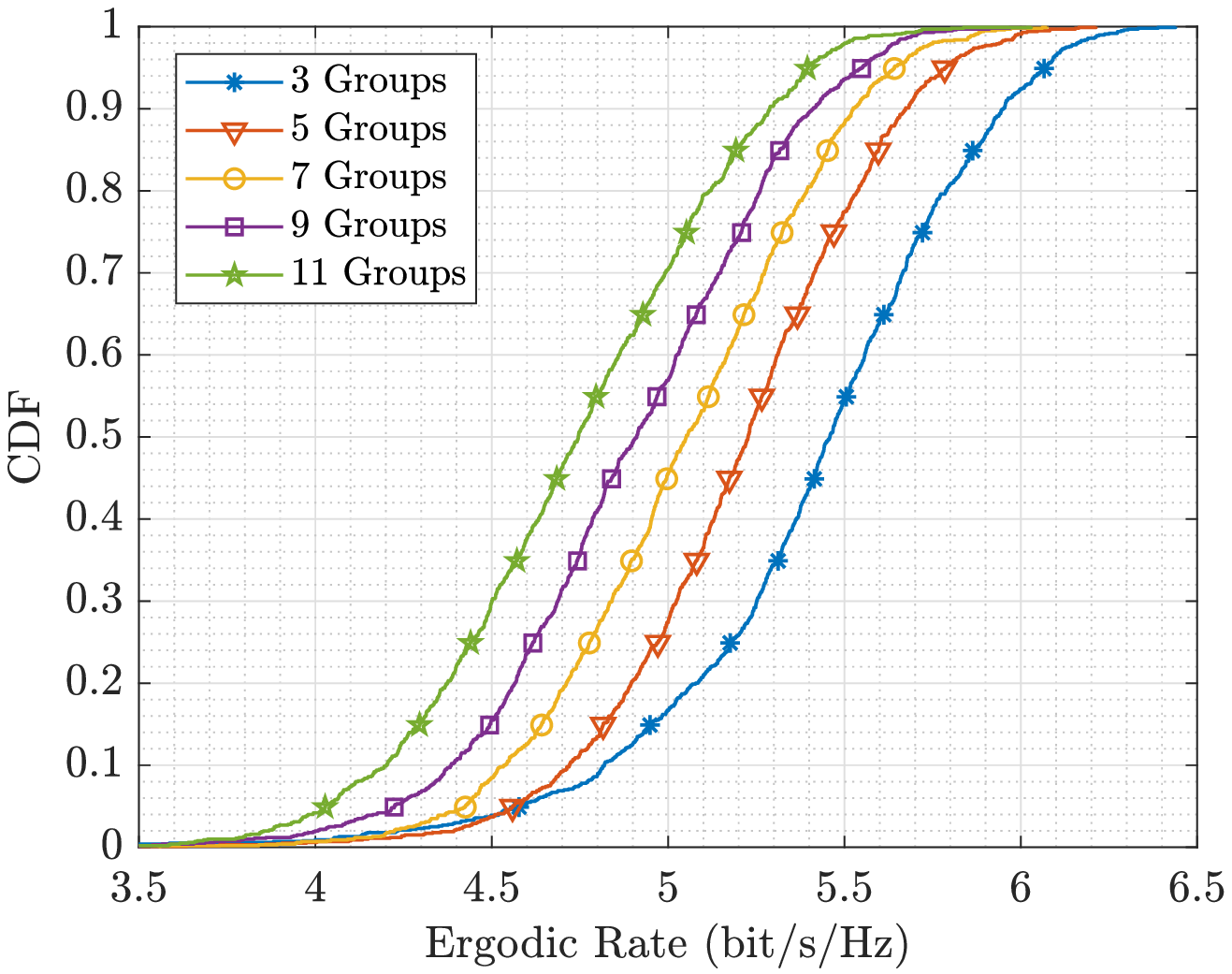}
    \caption{CDF of ergodic rates vs. number of groups in the NLoS case.}
    \label{fig:group_NLoS}
\end{figure}

We further consider how the SNR boost by blind beamforming scales with the number of REs $N$. The receiver position is now fixed at the origin point as shown in Fig. \ref{fig:simulation_loc}. Fig.~\ref {fig:N2_Rician} compares the SNR boost versus $N$ performance in the LoS case; we test beam training as well as CSM with different $T$ values. In particular, we remark that CSM with $T\rightarrow\infty$ is equivalent to the perfect CSI method. As shown in Fig.~\ref {fig:N2_Rician}, the SNR boost brought by beam training is approximately linear in $N$, while the rest algorithms yield faster growths of SNR boost in $N$---which are quasi-quadratic. Actually, with $T=KN^2$, CSM can almost reach its ideal status with infinitely many samples. Moreover, we can make a similar observation about the NLoS case from Fig.~\ref{fig:N2_Rayleigh}. We would like to take a closer look at the effect of different group schemes on the performance of the proposed GCSM algorithm. As shown in Fig.~\ref{fig:group_LoS} and Fig.~\ref{fig:group_NLoS}, the performance of GCSM degrades as the number of groups increases.

\begin{figure}[t]
    \centering
    \includegraphics[width=9cm]{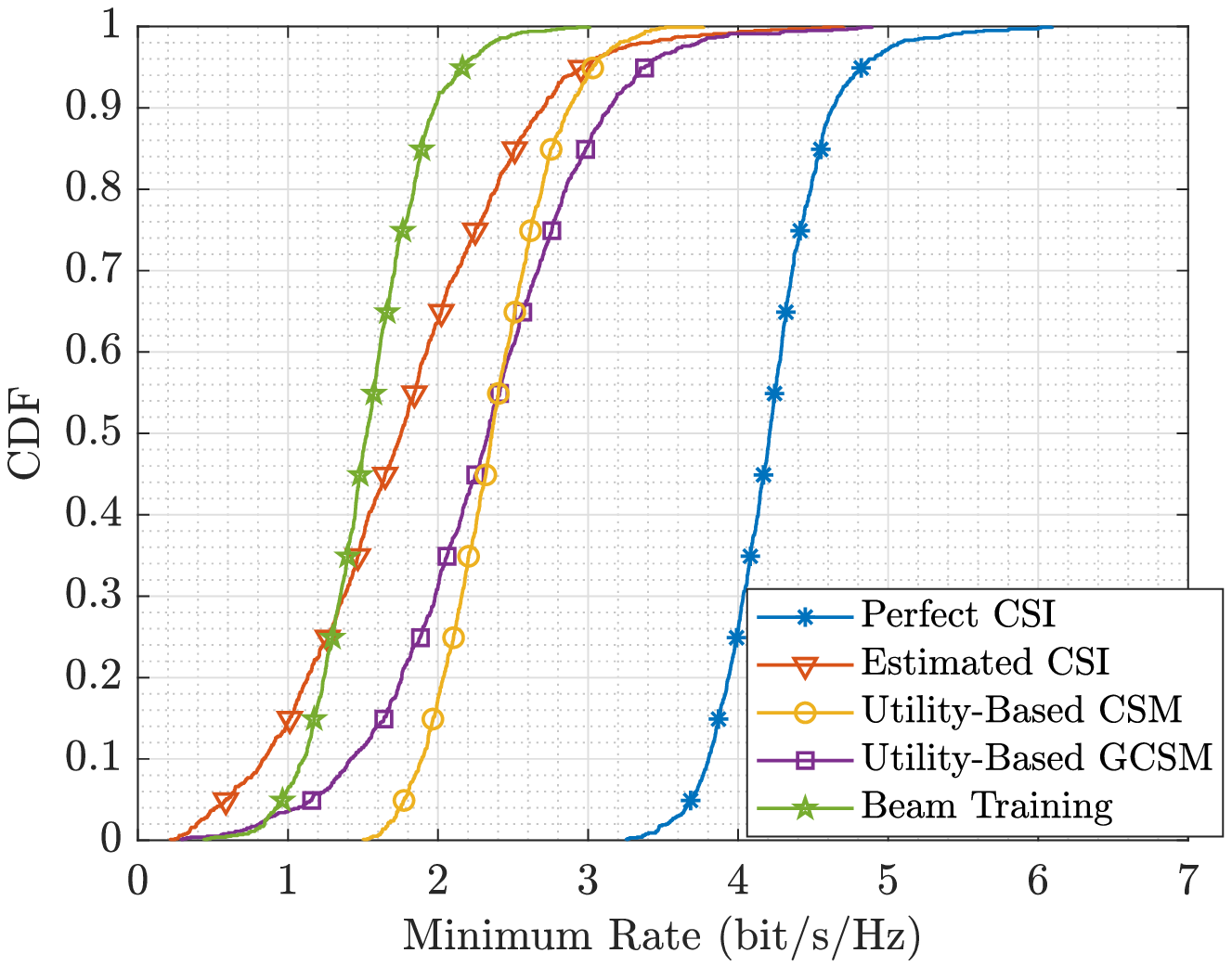}
    \caption{CDF of minimum rate in the LoS broadcast network.}
    \label{fig:CDF_LoS_multi}
\end{figure}

\begin{figure}[t]
    \centering
    \includegraphics[width=9cm]{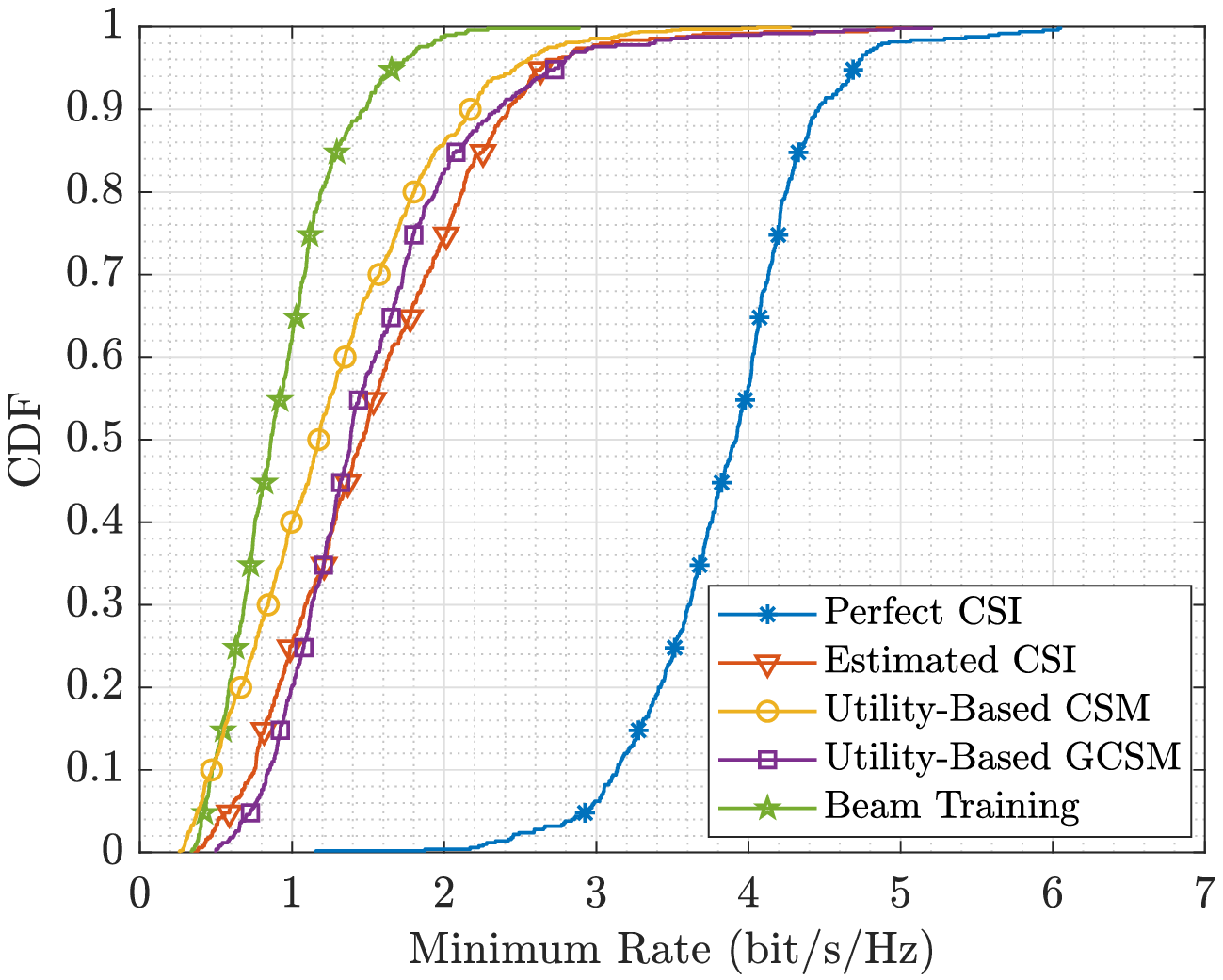}
    \caption{CDF of minimum rate in the NLoS broadcast network.}
    \label{fig:CDF_NLoS_multi}
\end{figure}

% \begin{figure}[t]
%     \centering
%     \includegraphics[width=9cm]{fig/double_CDF.eps}
%     \caption{Cumulative distributions of ergodic rate in the double-IS system.}
%     \label{fig:CDF_double}
% \end{figure}

% \begin{figure}[t]
%     \centering
%     \includegraphics[width=9cm]{fig/boost_vs_N_double.eps}
%     \caption{SNR boost versus $N$ in the double-IS system.}
%     \label{fig:quartic_boost_double}
% \end{figure}

Finally, we compare the different algorithms in a broadcast network. The locations of $M=5$ users are randomly generated within the shaded region shown in Fig.~\ref{fig:simulation_loc}. For both perfect CSI and estimated CSI, SDR is used to optimize the phase shift array $\bm \theta$. As shown in Fig.~\ref{fig:CDF_LoS_multi}, the proposed utility-based GCSM method outperforms estimated CSI and beam training significantly in terms of the minimum ergodic data rates when all direct channels are LoS. For instance, utility-based GCSM improves upon beam training by about 52\% and upon estimated CSI by about 33\% at the 50th percentile. Notice that utility-based GCSM yields a similar performance as utility-based CSM, indicating that it can also work well in the LoS case.

But when all direct channels are NLoS, the performance of all methods becomes worse. Observe that utility-based GCSM now outperforms the utility-based CSM. Observe also that the utility-based GCSM and the channel estimation based method yield similar performance in the NLoS case, whereas the former outperforms the latter significantly in the LoS case as formerly shown in Fig.~\ref{fig:CDF_LoS_multi}.
However, we argue that the proposed blind beamforming scheme can still be preferable to the channel estimation method even if they yield similar optimization performance, for two reasons. First, the computation burden of the proposed method is much smaller, as shown in Table \ref{tab:run_time}. Second, we remark that the channel estimation approach may not fit in the current network protocol and hardware. On the network protocol side, the channel estimation methods require the third-party intelligent surface to read the received symbol $Y\in\mathbb C$ from the communication chip of the receiver device, but this is not permitted in the current network protocol. On the hardware side, the existing intelligent surface prototype machines only support low-resolution phase shifting, e.g., $0$ or $\pi$ on each reflected element, but the channel estimation methods in the literature mostly require far more complicated phase shift settings, e.g., phase shifting according to the DFT matrix. As such, the existing prototypes of intelligent surface \cite{Arun2020RFocus, pei2021ris, tran2020demonstration, staat2022irshield} seldom adopt channel estimation.

% However, we argue that the utility-based GCSM is still much preferable to the estimated CSI due to the following reasons. First, the computational complexity of utility-based GCSM is $O(N(TM+K))$, which is significantly lower than the $O(N^{3.5})$ complexity of estimated CSI. The above complexity analysis is consistent with the running time comparison as shown in Table \ref{tab:run_time}.
% Second, the channel estimation in estimated CSI requires setting each phase shift according to the DFT matrix \cite{zheng2019intelligent}, which may violate the discrete phase shift constraint. For instance, it does not work when each phase shift is restricted to $\{0,\pi\}$.

\begin{table}[t]
\small
\renewcommand{\arraystretch}{1.3}
\centering
\caption{\small Running Time in the Broadcast Case}
\begin{tabular}{lr}
\hline
Method        & Running Time (second) \\ \hline
Perfect CSI & 1.29           \\
Estimated CSI & 1.90           \\
Utility-based CSM          & 0.11           \\
Utility-based GCSM         & 0.16           \\
Beam Training & 0.09           \\\hline
\end{tabular}
\label{tab:run_time}
\end{table}

% Finally, we validate the performance of blind beamforming in a double-IS system. The two ISs are placed as in Fig. \ref{fig:simulation_loc}. Notice that the LoS direct channel condition always holds in this case, so the sequential CSM method is used here.
% First, Fig.~\ref{fig:CDF_double} compares the cumulative distributions of ergodic rate of the different methods. Observe that the sequential CSM has a huge advantage over RMS and estimated CSI. For instance, in the 50th percentile, the sequential CSM increases the ergodic rate of RMS by about 90\% and increases that of the estimated CSI by about 40\%. Observe also that the sequential CSM can bring further gain if many more samples are provided---it will ultimately converge to the perfect CSI when $T\rightarrow\infty$. Moreover, as shown in Fig.~\ref{fig:quartic_boost_double}, the growth curve of the SNR boost versus $N$ resembles that in the single-IS case, only that the gap between blind beamforming and RMS now becomes even larger. The above result agrees with the preceding analysis in Proposition \ref{prop:SCSM_bound} that the SNR boost by the sequential CSM grows exponentially with the number of ISs.

\section{Conclusion}
\label{sec:Conclusion}
This work advocates a blind beamforming approach to the phase shift optimization problem of IS without channel acquisition. Although blind beamforming has been considered in the recent works \cite{Arun2020RFocus,ren2022configuring}, their discussions are limited to fixed channels, while this work extends blind beamforming to fading channels. More importantly, there is a subtle issue in \cite{Arun2020RFocus,ren2022configuring} that can completely jeopardize the existing RFocus algorithm and CSM algorithm when they are applied to the NLoS transmission case.
To address this issue, we suggest dividing all the REs into three groups and performing blind beamforming for them in an adaptive fashion. Furthermore, we extend the proposed algorithm to multiple users in a broadcast network. All the above results are numerically verified in the field tests and simulations. The proposed adaptive blind beamforming algorithm enables a fast configuration of IS without modifying the current network protocol, while not compromising the performance gain of the deployed IS.

\bibliographystyle{IEEEtran}     
\bibliography{IEEEabrv,Ref}

\begin{thebibliography}{10}
\providecommand{\url}[1]{#1}
\csname url@samestyle\endcsname
\providecommand{\newblock}{\relax}
\providecommand{\bibinfo}[2]{#2}
\providecommand{\BIBentrySTDinterwordspacing}{\spaceskip=0pt\relax}
\providecommand{\BIBentryALTinterwordstretchfactor}{4}
\providecommand{\BIBentryALTinterwordspacing}{\spaceskip=\fontdimen2\font plus
\BIBentryALTinterwordstretchfactor\fontdimen3\font minus
  \fontdimen4\font\relax}
\providecommand{\BIBforeignlanguage}[2]{{%
\expandafter\ifx\csname l@#1\endcsname\relax
\typeout{** WARNING: IEEEtran.bst: No hyphenation pattern has been}%
\typeout{** loaded for the language `#1'. Using the pattern for}%
\typeout{** the default language instead.}%
\else
\language=\csname l@#1\endcsname
\fi
#2}}
\providecommand{\BIBdecl}{\relax}
\BIBdecl

\bibitem{li2019towards}
Z.~Li, Y.~Xie, L.~Shangguan, R.~I. Zelaya, J.~Gummeson, W.~Hu, and K.~Jamieson,
  ``Towards programming the radio environment with large arrays of inexpensive
  antennas,'' in \emph{{USENIX} Symp. Netw. Sys. Design Implementation
  ({NSDI})}, Feb. 2019, pp. 285--300.

\bibitem{Arun2020RFocus}
V.~Arun and H.~Balakrishnan, ``{RFocus:} beamforming using thousands of passive
  antennas,'' in \emph{{USENIX} Symp. Netw. Sys. Design Implementation
  ({NSDI})}, Feb. 2020, pp. 1047--1061.

\bibitem{han2017enhancing}
S.~Han and K.~G. Shin, ``Enhancing wireless performance using reflectors,'' in
  \emph{Proc {IEEE} Int. Conf. Comput. Commun. (INFOCOM)}, May 2017.

\bibitem{xiong2017customizing}
X.~Xiong, J.~Chan, E.~Yu, N.~Kumari, A.~A. Sani, C.~Zheng, and X.~Zhou,
  ``Customizing indoor wireless coverage via 3{D}-fabricated reflectors,'' in
  \emph{ACM International Conference on Systems for Energy-Efficient Built
  Environments}, Nov. 2017.

\bibitem{li2024secure}
B.~Li, J.~Liao, W.~Wu, and Y.~Li, ``Intelligent reflecting surface assisted
  secure computation of wireless powered {MEC} system,'' \emph{{IEEE} Trans.
  Mobile Comput.}, vol.~23, no.~4, pp. 3048--3059, Apr. 2024.

\bibitem{Zargari2023energy}
S.~Zargari, C.~Tellambura, and S.~Herath, ``Energy-efficient hybrid offloading
  for backscatter-assisted wirelessly powered {MEC} with reconfigurable
  intelligent surfaces,'' \emph{{IEEE} Trans. Mobile Comput.}, vol.~22, no.~9,
  pp. 5262--5279, Sep. 2023.

\bibitem{shi2022intelligent}
W.~Shi, W.~Xu, X.~You, C.~Zhao, and K.~Wei, ``Intelligent reflection enabling
  technologies for integrated and green internet-of-everything beyond 5{G}:
  Communication, sensing, and security,'' \emph{{IEEE} Wireless Commun.},
  vol.~30, no.~2, pp. 147--154, Apr. 2023.

\bibitem{wu2024intelligent}
Q.~Wu, B.~Zheng, C.~You, L.~Zhu, K.~Shen, X.~Shao, W.~Mei, B.~Di, H.~Zhang,
  E.~Basar, L.~Song, M.~D. Renzo, Z.-Q. Luo, and R.~Zhang, ``Intelligent
  surfaces empowered wireless network: Recent advances and the road to 6{G},''
  \emph{Proc. {IEEE}}, 2024, to be published.

\bibitem{luo2010semidefinite}
Z.-Q. Luo, W.-K. Ma, A.~M.-C. So, Y.~Ye, and S.~Zhang, ``Semidefinite
  relaxation of quadratic optimization problems,'' \emph{{IEEE} Signal Process.
  Mag.}, vol.~27, no.~3, pp. 20--34, Apr. 2010.

\bibitem{wu2019intelligent}
Q.~Wu and R.~Zhang, ``Intelligent reflecting surface enhanced wireless network
  via joint active and passive beamforming,'' \emph{{IEEE} Trans. Wireless
  Commun.}, vol.~18, no.~11, pp. 5394--5409, Nov. 2019.

\bibitem{zheng2021double}
B.~Zheng, C.~You, and R.~Zhang, ``Double-{IRS} assisted multi-user {MIMO}:
  Cooperative passive beamforming design,'' \emph{{IEEE} Trans. Wireless
  Commun.}, vol.~20, no.~7, pp. 4513--4526, Jul. 2021.

\bibitem{zhou2020robust}
G.~Zhou, C.~Pan, H.~Ren, K.~Wang, M.~Di~Renzo, and A.~Nallanathan, ``Robust
  beamforming design for intelligent reflecting surface aided {MISO}
  communication systems,'' \emph{{IEEE} Wireless Commun. Lett.}, vol.~9,
  no.~10, pp. 1658--1662, Jun. 2020.

\bibitem{zeng2020sum}
M.~Zeng, X.~Li, G.~Li, W.~Hao, and O.~A. Dobre, ``Sum rate maximization for
  {IRS}-assisted uplink {NOMA},'' \emph{{IEEE} Commun. Lett.}, vol.~25, no.~1,
  pp. 234--238, Jan. 2020.

\bibitem{xie2020max}
H.~Xie, J.~Xu, and Y.-F. Liu, ``Max-min fairness in {IRS}-aided multi-cell
  {MISO} systems with joint transmit and reflective beamforming,'' \emph{{IEEE}
  Trans. Wireless Commun.}, vol.~20, no.~2, pp. 1379--1393, Feb. 2020.

\bibitem{huang2020decentralized}
S.~Huang, Y.~Ye, M.~Xiao, H.~V. Poor, and M.~Skoglund, ``Decentralized
  beamforming design for intelligent reflecting surface-enhanced cell-free
  networks,'' \emph{{IEEE} Wireless Commun. Lett.}, vol.~10, no.~3, pp.
  673--677, Jun. 2020.

\bibitem{yao2023superimposed}
J.~Yao, J.~Xu, W.~Xu, C.~Yuen, and X.~You, ``Superimposed {RIS}-phase
  modulation for {MIMO} communications: A novel paradigm of information
  transfer,'' \emph{{IEEE} Trans. Wireless Commun.}, vol.~23, no.~4, pp.
  2978--2993, Apr. 2024.

\bibitem{shen2018fractional_p1}
K.~Shen and W.~Yu, ``Fractional programming for communication systems—part
  {I}: Power control and beamforming,'' \emph{{IEEE} Trans. Signal Process.},
  vol.~66, no.~10, pp. 2616--2630, May 2018.

\bibitem{shen2018fractional_p2}
------, ``Fractional programming for communication systems—part {II}: Uplink
  scheduling via matching,'' \emph{{IEEE} Trans. Signal Process.}, vol.~66,
  no.~10, pp. 2631--2644, May 2018.

\bibitem{feng2020physical}
K.~Feng, X.~Li, Y.~Han, S.~Jin, and Y.~Chen, ``Physical layer security
  enhancement exploiting intelligent reflecting surface,'' \emph{{IEEE} Commun.
  Lett.}, vol.~25, no.~3, pp. 734--738, Mar. 2020.

\bibitem{zhu2020power}
J.~Zhu, Y.~Huang, J.~Wang, K.~Navaie, and Z.~Ding, ``Power efficient
  {IRS}-assisted {NOMA},'' \emph{{IEEE} Trans. Commun.}, vol.~69, no.~2, pp.
  900--913, Feb. 2020.

\bibitem{shafique2020optimization}
T.~Shafique, H.~Tabassum, and E.~Hossain, ``Optimization of wireless relaying
  with flexible {UAV}-borne reflecting surfaces,'' \emph{{IEEE} Trans.
  Commun.}, vol.~69, no.~1, pp. 309--325, Jan. 2020.

\bibitem{zhang2022active}
Z.~Zhang, L.~Dai, X.~Chen, C.~Liu, F.~Yang, R.~Schober, and H.~V. Poor,
  ``Active {RIS} vs. passive {RIS}: Which will prevail in {6G}?'' \emph{{IEEE}
  Trans. Commun.}, vol.~71, no.~3, pp. 1707--1725, Mar. 2022.

\bibitem{zhang2021joint}
Z.~Zhang and L.~Dai, ``A joint precoding framework for wideband reconfigurable
  intelligent surface-aided cell-free network,'' \emph{{IEEE} Trans. Signal
  Process.}, vol.~69, pp. 4085--4101, Jun. 2021.

\bibitem{palomar2016majorization}
Y.~Sun, P.~Babu, and D.~P. Palomar, ``Majorization-minimization algorithms in
  signal processing, communications, and machine learning,'' \emph{{IEEE}
  Trans. Signal Process.}, vol.~65, no.~3, pp. 794--816, Feb. 2016.

\bibitem{huang2019reconfigurable}
C.~Huang, A.~Zappone, G.~C. Alexandropoulos, M.~Debbah, and C.~Yuen,
  ``Reconfigurable intelligent surfaces for energy efficiency in wireless
  communication,'' \emph{{IEEE} Trans. Wireless Commun.}, vol.~18, no.~8, pp.
  4157--4170, Aug. 2019.

\bibitem{shen2020beamforming}
H.~Shen, W.~Xu, S.~Gong, C.~Zhao, and D.~W.~K. Ng, ``Beamforming optimization
  for {IRS}-aided communications with transceiver hardware impairments,''
  \emph{{IEEE} Trans. Commun.}, vol.~69, no.~2, pp. 1214--1227, Feb. 2020.

\bibitem{yu2024energy}
X.~Yu, G.~Wang, X.~Huang, K.~Wang, W.~Xu, and Y.~Rui, ``Energy efficient
  resource allocation for uplink {RIS}-aided millimeter-wave networks with
  {NOMA},'' \emph{{IEEE} Trans. Mobile Comput.}, vol.~23, no.~1, pp. 423--436,
  Jan. 2024.

\bibitem{yao2023robust}
J.~Yao, J.~Xu, W.~Xu, D.~W.~K. Ng, C.~Yuen, and X.~You, ``Robust beamforming
  design for {RIS}-aided cell-free systems with {CSI} uncertainties and
  capacity-limited backhaul,'' \emph{{IEEE} Trans. Commun.}, vol.~71, no.~8,
  pp. 4636--4649, Aug. 2023.

\bibitem{lai2024efficient}
W.~Lai, Z.~Wu, Y.~Feng, K.~Shen, and Y.-F. Liu, ``An efficient convex-hull
  relaxation based algorithm for multi-user discrete passive beamforming,''
  \emph{{IEEE} Signal Process. Lett.}, 2024, to be published.

\bibitem{pei2021ris}
X.~Pei, H.~Yin, L.~Tan, L.~Cao, Z.~Li, K.~Wang, K.~Zhang, and E.~Bj{\"o}rnson,
  ``{RIS}-aided wireless communications: Prototyping, adaptive beamforming, and
  indoor/outdoor field trials,'' \emph{{IEEE} Trans. Commun.}, vol.~69, no.~12,
  pp. 8627--8640, Dec. 2021.

\bibitem{tran2020demonstration}
N.~M. Tran, M.~M. Amri, D.~S. Kang, J.~H. Park, M.~H. Lee, D.~I. Kim, and K.~W.
  Choi, ``Demonstration of reconfigurable metasurface for wireless
  communications,'' in \emph{IEEE Wireless Commun. Netw. Conf. Workshops (WCNC
  Workshops)}, Apr. 2020.

\bibitem{staat2022irshield}
P.~Staat, S.~Mulzer, S.~Roth, V.~Moonsamy, M.~Heinrichs, R.~Kronberger,
  A.~Sezgin, and C.~Paar, ``{IRS}hield: A countermeasure against adversarial
  physical-layer wireless sensing,'' in \emph{IEEE Symp. Secur. Priv. (SP)},
  May 2022.

\bibitem{ren2022configuring}
S.~Ren, K.~Shen, Y.~Zhang, X.~Li, X.~Chen, and Z.-Q. Luo, ``Configuring
  intelligent reflecting surface with performance guarantees: Blind
  beamforming,'' \emph{{IEEE} Trans. Wireless Commun.}, vol.~22, no.~5, pp.
  3355--3370, May 2023.

\bibitem{zappone2021intelligent}
Q.-U.-A. Nadeem, A.~Zappone, and A.~Chaaban, ``Intelligent reflecting surface
  enabled random rotations scheme for the {MISO} broadcast channel,''
  \emph{{IEEE} Trans. Wireless Commun.}, vol.~20, no.~8, pp. 5226--5242, Aug.
  2021.

\bibitem{psomas2021low}
C.~Psomas and I.~Krikidis, ``Low-complexity random rotation-based schemes for
  intelligent reflecting surfaces,'' \emph{{IEEE} Trans. Wireless Commun.},
  vol.~20, no.~8, pp. 5212--5225, Aug. 2021.

\bibitem{cai2023toward}
X.~Cai, C.~Huang, E.~Basar, W.~Xu, L.~Wang, M.~Di~Renzo, and C.~Yuen, ``Toward
  {RIS}-aided non-coherent communications: A joint index keying {M}-ary
  differential chaos shift keying system,'' \emph{{IEEE} Trans. Wireless
  Commun.}, vol.~22, no.~12, pp. 9045--9062, Dec. 2023.

\bibitem{you2020fast}
C.~You, B.~Zheng, and R.~Zhang, ``Fast beam training for {IRS}-assisted
  multiuser communications,'' \emph{{IEEE} Wireless Commun. Lett.}, vol.~9,
  no.~11, pp. 1845--1849, Nov. 2020.

\bibitem{ning2021terahertz}
B.~Ning, Z.~Chen, W.~Chen, Y.~Du, and J.~Fang, ``Terahertz multi-user massive
  {MIMO} with intelligent reflecting surface: Beam training and hybrid
  beamforming,'' \emph{{IEEE} Trans. Veh. Technol.}, vol.~70, no.~2, pp.
  1376--1393, Feb. 2021.

\bibitem{wang2022fast}
P.~Wang, J.~Fang, W.~Zhang, and H.~Li, ``Fast beam training and alignment for
  {IRS}-assisted millimeter wave/terahertz systems,'' \emph{{IEEE} Trans.
  Wireless Commun.}, vol.~21, no.~4, pp. 2710--2724, Apr. 2022.

\bibitem{wang2021jointbeam}
W.~Wang and W.~Zhang, ``Joint beam training and positioning for intelligent
  reflecting surfaces assisted millimeter wave communications,'' \emph{{IEEE}
  Trans. Wireless Commun.}, vol.~20, no.~10, pp. 6282--6297, Oct. 2021.

\bibitem{Xu2024Coordinating}
F.~Xu, J.~Yao, W.~Lai, K.~Shen, X.~Li, X.~Chen, and Z.-Q. Luo, ``Coordinating
  multiple intelligent reflecting surfaces without channel information,''
  \emph{{IEEE} Trans. Signal Process.}, vol.~72, pp. 31--46, 2024.

\bibitem{xu2024blind}
------, ``Blind beamforming for coverage enhancement with intelligent
  reflecting surface,'' \emph{{IEEE} Trans. Wireless Commun.}, 2024, to be
  published.

\bibitem{xu2023reconfiguring}
J.~Xu, C.~Yuen, C.~Huang, N.~Ul~Hassan, G.~C. Alexandropoulos, M.~Di~Renzo, and
  M.~Debbah, ``Reconfiguring wireless environments via intelligent surfaces for
  6{G}: Reflection, modulation, and security,'' \emph{Science China Information
  Sciences}, vol.~66, no.~3, pp. 130\,304:1--130\,304:20, Feb. 2023.

\bibitem{2022Xuwy}
W.~Xu, L.~Gan, and C.~Huang, ``A robust deep learning-based beamforming design
  for {RIS}-assisted multiuser {MISO} communications with practical
  constraints,'' \emph{{IEEE} Trans. on Cogn. Commun. Netw.}, vol.~8, no.~2,
  pp. 694--706, Jun. 2022.

\bibitem{2020HuangcwJSAC}
C.~Huang, R.~Mo, and C.~Yuen, ``Reconfigurable intelligent surface assisted
  multiuser {MISO} systems exploiting deep reinforcement learning,''
  \emph{{IEEE} J. Sel. Areas Commun.}, vol.~38, no.~8, pp. 1839--1850, Aug.
  2020.

\bibitem{2021JiangtJSAC}
T.~Jiang, H.~V. Cheng, and W.~Yu, ``Learning to reflect and to beamform for
  intelligent reflecting surface with implicit channel estimation,''
  \emph{{IEEE} J. Sel. Areas Commun.}, vol.~39, no.~7, pp. 1931--1945, Jul.
  2021.

\bibitem{zhang2021large}
J.~Zhang, J.~Liu, S.~Ma, C.-K. Wen, and S.~Jin, ``Large system achievable rate
  analysis of {RIS}-assisted {MIMO} wireless communication with statistical
  {CSIT},'' \emph{{IEEE} Trans. Wireless Commun.}, vol.~20, no.~9, pp.
  5572--5585, Sep. 2021.

\bibitem{xu2021sum}
K.~Xu, J.~Zhang, X.~Yang, S.~Ma, and G.~Yang, ``On the sum-rate of
  {RIS}-assisted {MIMO} multiple-access channels over spatially correlated
  {R}ician fading,'' \emph{{IEEE} Trans. Commun.}, vol.~69, no.~12, pp.
  8228--8241, Dec. 2021.

\bibitem{zhang2022sum}
H.~Zhang, S.~Ma, Z.~Shi, X.~Zhao, and G.~Yang, ``Sum-rate maximization of
  {RIS}-aided multi-user {MIMO} systems with statistical {CSI},'' \emph{{IEEE}
  Trans. Wireless Commun.}, vol.~22, no.~7, pp. 4788--4801, Jul. 2023.

\bibitem{luo2021reconfigurable}
C.~Luo, X.~Li, S.~Jin, and Y.~Chen, ``Reconfigurable intelligent
  surface-assisted multi-cell {MISO} communication systems exploiting
  statistical {CSI},'' \emph{{IEEE} Wireless Commun. Lett.}, vol.~10, no.~10,
  pp. 2313--2317, Oct. 2021.

\bibitem{jiang2022ris}
L.~Jiang, C.~Luo, X.~Li, M.~Matthaiou, and S.~Jin, ``{RIS}-assisted downlink
  multi-cell communication using statistical {CSI},'' in \emph{2022
  International Symposium on Wireless Communication Systems (ISWCS)}, Oct.
  2022.

\bibitem{jia2020analysis}
Y.~Jia, C.~Ye, and Y.~Cui, ``Analysis and optimization of an intelligent
  reflecting surface-assisted system with interference,'' \emph{{IEEE} Trans.
  Wireless Commun.}, vol.~19, no.~12, pp. 8068--8082, Dec. 2020.

\bibitem{peng2021analysis}
Z.~Peng, T.~Li, C.~Pan, H.~Ren, W.~Xu, and M.~Di~Renzo, ``Analysis and
  optimization for {RIS}-aided multi-pair communications relying on statistical
  {CSI},'' \emph{{IEEE} Trans. Veh. Technol.}, vol.~70, no.~4, pp. 3897--3901,
  Apr. 2021.

\bibitem{dai2021statistical}
J.~Dai, F.~Zhu, C.~Pan, H.~Ren, and K.~Wang, ``Statistical {CSI}-based
  transmission design for reconfigurable intelligent surface-aided massive
  {MIMO} systems with hardware impairments,'' \emph{{IEEE} Wireless Commun.
  Lett.}, vol.~11, no.~1, pp. 38--42, Jan. 2021.

\bibitem{zhi2022power}
K.~Zhi, C.~Pan, H.~Ren, and K.~Wang, ``Power scaling law analysis and phase
  shift optimization of {RIS}-aided massive {MIMO} systems with statistical
  {CSI},'' \emph{{IEEE} Trans. Commun.}, vol.~70, no.~5, pp. 3558--3574, May
  2022.

\bibitem{cao2022two}
Y.~Cao, T.~Lv, and W.~Ni, ``Two-timescale optimization for intelligent
  reflecting surface-assisted {MIMO} transmission in fast-changing channels,''
  \emph{{IEEE} Trans. Wireless Commun.}, vol.~21, no.~12, pp. 10\,424--10\,437,
  Dec. 2022.

\bibitem{shekhar2022instantaneous}
S.~Shekhar, A.~Subhash, T.~Kella, and S.~Kalyani, ``Instantaneous channel
  oblivious phase shift design for an {IRS}-assisted {SIMO} system with
  quantized phase shift,'' 2022, [Online]. Available:
  https://arxiv.org/abs/2211.03317.

\bibitem{eskandari2022statistical}
M.~Eskandari, H.~Zhu, A.~Shojaeifard, and J.~Wang, ``Statistical {CSI}-based
  beamforming for {RIS}-aided multiuser {MISO} systems using deep reinforcement
  learning,'' 2022, [Online]. Available: https://arxiv.org/abs/2209.09856.

\bibitem{ren2022long}
H.~Ren, C.~Pan, L.~Wang, W.~Liu, Z.~Kou, and K.~Wang, ``Long-term {CSI}-based
  design for {RIS}-aided multiuser {MISO} systems exploiting deep reinforcement
  learning,'' \emph{{IEEE} Commun. Lett.}, vol.~26, no.~3, pp. 567--571, Mar.
  2022.

\bibitem{zhongze_2022_learning}
Z.~Zhang, T.~Jiang, and W.~Yu, ``Learning based user scheduling in
  reconfigurable intelligent surface assisted multiuser downlink,''
  \emph{{IEEE} J. Sel. Topics Signal Process.}, vol.~16, no.~5, pp. 1026--1039,
  Aug. 2022.

\bibitem{guo2020intelligent}
H.~Guo, Y.-C. Liang, and S.~Xiao, ``Intelligent reflecting surface
  configuration with historical channel observations,'' \emph{{IEEE} Wireless
  Commun. Lett.}, vol.~9, no.~11, pp. 1821--1824, Nov. 2020.

\bibitem{goldsmith2005wireless}
A.~Goldsmith, \emph{Wireless communications}.\hskip 1em plus 0.5em minus
  0.4em\relax Cambridge university press, 2005.

\bibitem{han2019large}
Y.~Han, W.~Tang, S.~Jin, C.-K. Wen, and X.~Ma, ``Large intelligent
  surface-assisted wireless communication exploiting statistical {CSI},''
  \emph{{IEEE} Trans. Veh. Technol.}, vol.~68, no.~8, pp. 8238--8242, Aug.
  2019.

\bibitem{gan2021ris}
X.~Gan, C.~Zhong, C.~Huang, and Z.~Zhang, ``{RIS}-assisted multi-user {MISO}
  communications exploiting statistical {CSI},'' \emph{{IEEE} Trans. Commun.},
  vol.~69, no.~10, pp. 6781--6792, Oct. 2021.

\bibitem{wang2021joint}
J.~Wang, H.~Wang, Y.~Han, S.~Jin, and X.~Li, ``Joint transmit beamforming and
  phase shift design for reconfigurable intelligent surface assisted {MIMO}
  systems,'' \emph{{IEEE} Trans. on Cogn. Commun. Netw.}, vol.~7, no.~2, pp.
  354--368, Jun. 2021.

\bibitem{zhang2022configuring}
Y.~Zhang, K.~Shen, S.~Ren, X.~Li, X.~Chen, and Z.-Q. Luo, ``Configuring
  intelligent reflecting surface with performance guarantees: Optimal
  beamforming,'' \emph{{IEEE} J. Sel. Topics Signal Process.}, vol.~16, no.~5,
  pp. 967--979, Aug. 2022.

\bibitem{ren2022linear}
S.~Ren, K.~Shen, X.~Li, X.~Chen, and Z.-Q. Luo, ``A linear time algorithm for
  the optimal discrete {IRS} beamforming,'' \emph{{IEEE} Wireless Commun.
  Lett.}, vol.~12, no.~3, pp. 496--500, Mar. 2022.

\bibitem{yan2023passive}
G.~Yan, L.~Zhu, and R.~Zhang, ``Passive reflection optimization for {IRS}-aided
  multicast beamforming with discrete phase shifts,'' \emph{{IEEE} Wireless
  Commun. Lett.}, vol.~12, no.~8, pp. 1424--1428, Aug. 2023.

\bibitem{jiang2021learning}
T.~Jiang, H.~V. Cheng, and W.~Yu, ``Learning to reflect and to beamform for
  intelligent reflecting surface with implicit channel estimation,''
  \emph{{IEEE} J. Sel. Areas Commun.}, vol.~39, no.~7, pp. 1931--1945, Jul.
  2021.

\bibitem{zheng2019intelligent}
B.~Zheng and R.~Zhang, ``Intelligent reflecting surface-enhanced {OFDM}:
  Channel estimation and reflection optimization,'' \emph{{IEEE} Wireless
  Commun. Lett.}, vol.~9, no.~4, pp. 518--522, Apr. 2019.

\end{thebibliography}

\begin{IEEEbiographynophoto}{Wenhai Lai}(Graduate Student Member, IEEE) received the B.E. degree in information engineering from Beijing University of Posts and Telecommunications, in 2021. He is currently working toward the Ph.D. degree with the School of Science and Engineering, The Chinese University of Hong Kong, Shenzhen, China. His research interests include intelligent reflecting surface and reinforcement learning.
\end{IEEEbiographynophoto}

\begin{IEEEbiographynophoto}{Wenyu Wang}(Graduate Student Member, IEEE) received the B.E. degree in digital media technology from Huazhong University of Science and Technology in 2020. He is currently working toward the Ph.D degree with the School of Science and Engineering, The Chinese University of Hong Kong, Shenzhen, China. His research interests include intelligent reflecting surface and ultra-reliable and low-latency communication.
\end{IEEEbiographynophoto}

\begin{IEEEbiographynophoto}{Fan Xu}(Member, IEEE) received the B.S. degree in physics and the Ph.D degree in information and communication engineering from Shanghai Jiao Tong University, Shanghai, China, in 2016 and 2022, respectively. He received Huawei Scholarship in 2018 and was the outstanding graduate of Shanghai Jiao Tong University in 2022. From 2022 to 2024, he joined Peng Cheng Laboratory, Shenzhen, China, as a post-doctor. He is currently an assistant professor with the School of Electronic Information Engineering, Tongji University, Shanghai, China. His research interests include coded caching, distributed computing, intelligent reflecting surface, signal processing and optimization of 5G and beyond networks.
\end{IEEEbiographynophoto}

\begin{IEEEbiographynophoto}{Xin Li}
graduated from Xidian University and joined Huawei in 2008. He has rich experience in wireless channel modeling and wireless network performance modeling and optimization. Currently, he is a technical expert in Huawei's experience lab, focusing on future-oriented network technology research, including new technologies such as Intelligent Reflection Surface and Intelligent Transmission Surface, and their application in network structure optimization.
\end{IEEEbiographynophoto}

\begin{IEEEbiographynophoto}{Shaobo Niu} graduated from Xidian University and joined Huawei in 2010. He has rich experience in wireless network planning and optimization. Currently, he is a technical expert in Huawei's Algorithm Technical Research Dept, focusing on 5G wireless network simulation and optimization, including research on application of IRS in real network.
\end{IEEEbiographynophoto}

\begin{IEEEbiographynophoto}{Kaiming Shen}(Senior Member, IEEE) received the B.Eng. degree in information security and the B.Sc. degree in mathematics from Shanghai Jiao Tong University, China in 2011, and then the M.A.Sc. degree in electrical and computer engineering from the University of Toronto, Canada in 2013. After working at a tech startup in Ottawa for one year, he returned to the University of Toronto and received the Ph.D. degree in electrical and computer engineering in early 2020. Dr. Shen has been with the School of Science and Engineering at The Chinese University of Hong Kong (CUHK), Shenzhen, China as a tenure-track assistant professor since 2020. His research interests include optimization, wireless communications, information theory, and machine learning.
Dr. Shen received the IEEE Signal Processing Society Young Author Best Paper Award in 2021, the CUHK Teaching Achievement Award in 2023, and the Frontiers of Science Award in 2024. Dr. Shen currently serves as an Editor for IEEE Transactions on Wireless Communications.
\end{IEEEbiographynophoto}

\end{document}